\documentclass[showpacs,twocolumn,superscriptaddress,nofootinbib]{revtex4}

\usepackage{doi}%<----------
\usepackage{hyperref}
\hypersetup{
  colorlinks=true,        % false: boxed links; true: colored links
  linkcolor=blue,         % color of internal links
  citecolor=cyan,         % color of links to bibliography
}

\usepackage{graphicx}% Include figure files
\usepackage{dcolumn}% Align table columns on decimal point
\usepackage{bm}% bold math
\usepackage{color}

\begin{document}
%opening
\title{Shadow of rotating regular black holes}

\author{Ahmadjon Abdujabbarov}
\email{ahmadjon@astrin.uz}

\affiliation{Institute of Nuclear Physics,
 Ulughbek, Tashkent 100214, Uzbekistan}

\affiliation{Ulugh Beg Astronomical Institute, Astronomicheskaya 33,
 Tashkent 100052, Uzbekistan}
\affiliation{National University of Uzbekistan, Tashkent 100174, Uzbekistan}

\author{Muhammed Amir}
\email{amirctp12@gmail.com}

\affiliation{Centre for Theoretical Physics,
 Jamia Millia Islamia,  New Delhi 110025,
 India}

\author{Bobomurat Ahmedov}
\email{ahmedov@astrin.uz}

\affiliation{Institute of Nuclear Physics,
 Ulughbek, Tashkent 100214, Uzbekistan}

\affiliation{Ulugh Beg Astronomical Institute, Astronomicheskaya 33,
 Tashkent 100052, Uzbekistan}

\affiliation{National University of Uzbekistan, Tashkent 100174, Uzbekistan}

\author{Sushant G. Ghosh}
\email{sghosh2@jmi.ac.in}

\affiliation{Centre for Theoretical Physics,
 Jamia Millia Islamia,  New Delhi 110025,
 India}

\affiliation{Astrophysics and Cosmology Research Unit,
 School of Mathematics, Statistics and Computer Science,
 University of KwaZulu-Natal, Private Bag X54001,
 Durban 4000, South Africa}

\begin{abstract}
We study the shadows cast by the different types of rotating regular black holes viz. Ay\'{o}n-Beato-Garc\'{i}a {(ABG)}, Hayward, and Bardeen. These black holes have in addition to the total mass ($M$) and rotation parameter ($a$), different parameters as electric charge ($Q$), deviation parameter ($g$), and magnetic charge ($g_{*}$), respectively. Interestingly, the size of the shadow is affected by these parameters in addition to the rotation parameter. We found that the radius of the shadow in each case decreases monotonically and the distortion parameter increases when the value of these parameters increase. A comparison with the standard Kerr case is also investigated.
We have also studied the influence of the plasma environment around regular black holes to discuss its shadow. The presence of the plasma affects the apparent size of the regular black hole's shadow to be increased due to two effects (i) gravitational redshift of the photons and (ii) radial dependence of plasma density.

\end{abstract}

\maketitle

\section{Introduction}

As we know that the black holes are not visible objects, hence it is interesting to study the null geodesics around them. The photons can be considered to extract the information from the black holes. The way is provided by gravitational lensing by the black holes which has been demonstrated for the last few decades by several authors. Both the deflection of light and the change in apparent brightness of the radiating source by a gravitational field are known as a gravitational lens.
%
% The theory of gravitational lensing behaved very well for the weak field approximation case but the influence of the strong gravity appears only when the photon crosses black hole environment.
%
%
{In many cases the lensing in the weak gravitational field
approximation describes the scenario fully. However, in the case of
ultra-compact hypothetical objects like boson stars the occurrence
of the light rings are due to strong gravitational lensing (see,
e.g.~\cite{Cunha15}). }
Various authors have studied the gravitational lenses with a rotating black hole as a deflector \cite{Sereno03} and focused only on null geodesics motion at the equatorial plane. For example, the gravitational lensing by Kerr black hole is discussed in \cite{Bozza03,Vazquez04,Bozza2005,Bozza06a,Bozza07,Kraniotis11}. The apparent shape of the nonrotating black holes is a perfect circle while it is deformed for the rotating black holes due to the presence of spin \cite{Bardeen73,Chandrasekhar98}. The topic of gravitational lensing has been discussed by several authors with the expectation that the direct observation of black hole horizons will be possible in the near future \cite{Zakharov05,Chan15,Psaltis15}.

{To resolve the invisibility of the black hole there is Event Horizon Telescope (EHT)\footnote{http://www.eventhorizontelescope.org/} to achieve the angular resolution comparable to a black hole shadow.} A black hole casts a shadow, if it is in front of a distant bright object. The investigation of observing the black hole shadow is very interesting and useful tool for measuring the nature of astrophysical black holes. The observation of the shadow also provides a tentative way to find out the parameters of the black hole.

The shadow cast by the Schwarzschild black hole is first of all
discussed by the Synge \cite{Synge66} and Luminet \cite{Luminet79}.
Synge gave a formula to calculate the angular radius of the shadow.
Bardeen \cite{Bardeen73} was the first who studied the appearance of
the shadow cast by the Kerr black hole, the result can be seen in
Chandrasekhar's book \cite{Chandrasekhar98} {and in~\cite{Vries00}}.
It can be seen for Kerr black hole the shadow is no longer circular.
The shadow of the Kerr black hole or a Kerr naked singularity by
constructing two observables has been discussed by Hioki and Maeda
\cite{Hioki09}. Recently, some authors of this paper have developed
new coordinate independent formalism to describe the shadow of the
black holes~\cite{Abdujabbarov15}. This subject for other black
holes has been discussed by several authors, e.g. Kerr-Newman black
hole \cite{Takahashi05,Vries00}, Einstein-Maxwell-Dilaton-Axion
black hole \cite{Wei13}, Kerr-Taub-NUT black hole
\cite{Abdujabbarov13c}, rotating braneworld black hole
\cite{Amarilla12}, Kaluza-Klein rotating dilaton black hole
\cite{Amarilla13}, rotating non-Kerr black hole
\cite{Atamurotov13b}, Kerr-Newman-NUT black holes with cosmological
constant \cite{Grenzebach2014}. The subject of getting shadow has
also been extended for 5D rotating Myers-Perry black hole
\cite{Papnoi14}. {An example of a single black hole solution of
general relativity with multiple shadows has been shown, for the
first time, in \cite{Cunha15}.}

{\citet{Falcke00} have initiated that very long baseline
interferometry  radio interferometers with the advanced high spatial
resolution would be able to resolve the supermassive black hole
event horizon located at the centre of either the Milky Way or the
M87 galaxy in the submillimeter wavelength diapason (see, for the
furthermore details, \cite{Takahashi04,Johannsen10,Falcke13b}).}

In this paper our aim is to extend the discussion of black hole
shadow for  rotating regular black holes (e.g.
Ay\'{o}n-Beato-Garc\'{i}a, Hayward, and Bardeen), and to see the
effect of the parameters on the size of a shadow and on distortion
of a shadow. We also plan to study the influence of the plasma on the
optical properties of the regular black holes. Our recent paper has
been devoted to study the optical properties of the Kerr black
hole~\cite{Atamurotov15a}. Influence of plasma to the shadow of
static black holes has been considered in~\cite{Perlick15}. Optical
phenomena in the field of Braneworld Kerr black hole has been
studied in~\cite{Schee09}. The analyzation of the circular geodesics
around some regular black holes presented in~\cite{Stuchlik15}.
Optical and other properties of Kerr superspinars have been
considered in~\cite{Stuchlik10,Stuchlik12a,Stuchlik12b}.
Gravitational lensing effects have been widely studied in the
literature~\cite{Bisnovatyi2010,Tsupko12,Morozova13,Perlick15,Er14,Rogers15}.

The paper is organized as follows. In Sect.~\ref{sect2}, we study
apparent shape of the shadow of rotating Ay\'{o}n-Beato-Garc\'{i}a
{(ABG)} black hole and calculate the corresponding observable. We
discuss about the energy emission rate of ABG black in the
subsection of Sect.~\ref{sect2}.  In Sect.~\ref{sect3}, we study the
apparent shape of the shadow cast by the rotating Hayward and
Bardeen black holes and also see the behavior of the observable, and
in subsection we study the energy emission rate  for rotating
Hayward and Bardeen black holes. The Sect.~\ref{sect4} is devoted to
study the plasma influence on shadow of regular black holes. We
conclude our results in Sect.~\ref{sectconcl}.  We have fixed units
such that $G=c=1$.

\section{Rotating Ay\'{o}n-Beato-Garc\'{i}a black hole\label{sect2}}

We start with the rotating Ay\'{o}n-Beato-Garc\'{i}a (ABG) spacetime
which is a nonsingular exact black hole solution of Einstein field
equations coupled to a nonlinear electrodynamics and satisfying the
weak energy condition. This spacetime class was introduce
by Ay\'{o}n-Beato et al.
\cite{Ayon-Beato98,Ayon-Beato99,Ayon-Beato99a} and the rotating one
is discussed by Toshmatov et al.~\cite{Toshmatov14,Azreg14}. The
background metric of a rotating ABG
spacetime in the Boyer-Lindquist coordinates ($t$, $r$,
$\theta$, $\phi$) reads:
\begin{eqnarray}
\label{metric}
d{s}^2 &=&-f(r,\theta) dt^2+\frac{\Sigma}{\Delta} dr^2
\nonumber \\ &&
-2a\sin^2\theta(1-f(r,\theta))d\phi dt+\Sigma d\theta^2 \nonumber \\ &&
+ \sin^2\theta[\Sigma-a^2(f(r,\theta)-2)\sin^2\theta]d\phi^2,
\end{eqnarray}
where the metric function $f(r,\theta)$ is given by
\begin{equation}
\label{f}
f(r,\theta) = 1-\frac{2M r \sqrt{\Sigma}}{(\Sigma+Q^2)^{3/2}}+\frac{Q^2\Sigma}{(\Sigma+Q^2)^2},
\end{equation}
with
\begin{equation}
\label{delta}
\Delta = \Sigma f(r,\theta) + a^2\sin^2\theta, \;\;\;\; \Sigma = r^2+a^2 \cos^2 \theta,
\end{equation}
where $M$ is the mass, $a$ is rotation parameter, and $Q$ is the electric charge of the black hole. The stationary and axial-symmetric metric (\ref{metric}) contains four constants of motion which are the Lagrangian ($\mathcal{L}$), the energy ($E$), $z$-component of angular momentum ($L_{z}$), and the Carter constant ($\mathcal{K}$).

In order to discuss the black hole shadow, we need to calculate the geodesic equations of the photons for the metric (\ref{metric}). It is very difficult to separate the constants when we apply the Hamilton Jacobi formulation for the rotating ABG spacetime because the function $f(r,\theta)$ has very complicated form. Therefore, to resolve this problem, we consider an approximation in $\theta$, such that $\theta \approx \pi/2+\epsilon$, where $\epsilon$ is small angle.
{Note, that here we consider the near-equatorial plane orbits of the photons, however, unstable
photon circular orbits are not restricted necessarily to
the equatorial plane. This fact does not render the
calculations below, since in this paper our main aim is calculating the shadow of black hole by the observer at the infinity,  which can indeed be obtained using the above mentioned
approximation. Furthermore, as the observer is situated far away from the black hole, the photons will arrive near the
equatorial plane (see for the details the Subsection A below of this Section).}
In this case the trigonometric functions take the form as $\sin \theta =1$, $\cos \theta =\epsilon$ and the function $f(r)$ is given in simple form:
\begin{equation}
\label{fr}
f(r) = 1-\frac{2M r^2}{(r^2+Q^2)^{3/2}}+\frac{Q^2r^2}{(r^2+Q^2)^2}.
\end{equation}
We can easily get the following geodesic equations by solving equations $E=-p_{t}=-\partial \mathcal{L} / \partial \dot{t}$ and $L_{z}=p_{\phi}=\partial \mathcal{L} / \partial \dot{\phi}$, simultaneously
\begin{equation}
\label{abg1}
r^2\frac{d t}{d \sigma} =  a (L_{z}-aE) +  \frac{r^2 + a^2}{\Delta}\left[(r^2 + a^2)E-aL_{z}\right],
\end{equation}
%%%%%%%%%%%%%%
\begin{equation}
\label{abg2}
r^2\frac{d \phi}{d \sigma} = (L_{z}-aE) + \frac{a}{\Delta}\left((r^2 + a^2)E-aL_{z}\right),
\end{equation}
%%%%%%%%%%%%%%
where $\sigma$ is an affine parameter along the geodesics. Now, we can easily find out the remaining geodesic equations by using the Hamilton-Jacobi formulation. The corresponding Hamilton-Jacobi equation has the following form
\begin{equation}
\label{hje}
\frac{\partial S}{\partial \sigma} = -\frac{1}{2} g^{\mu\nu} \frac{\partial S}{\partial x^{\mu}} \frac{\partial S}{\partial x^{\nu}},
\end{equation}
where $S$ is Jacobi action. If we have a separable solution, then it takes the following form
\begin{equation}
\label{hja}
S = \frac{1}{2} m_0^2 \sigma -Et + L_{z} \phi + S_{r}(r) + S_{\epsilon}(\epsilon),
\end{equation}
where $ S_{r}(r) $ and $S_{\epsilon}(\epsilon)$ are functions of $r$ and $\epsilon$, respectively. Inserting Eq.~(\ref{hja}) into the Eq.~(\ref{hje}) and separating out the coefficients of $r$ and $\epsilon$ being equal to the Carter constant then we can easily get the geodesic equations in the following form
\begin{equation}
\label{abg3}
r^2 \frac{d r}{d \sigma} = \pm\sqrt{\mathcal{R}},
\end{equation}
\begin{equation}
\label{abg4}
r^2 \frac{d \epsilon}{d \sigma} = \pm \sqrt{\Theta},
\end{equation}
where
\begin{equation}
\label{R}
\mathcal{R}= \left[(r^2 + a^2)E-aL_{z}\right]^2-\Delta \left[\mathcal{K}+(L_{z}-aE)^2\right],
\end{equation}
\begin{equation}
\label{Th}
\Theta= \mathcal{K},
\end{equation}
{where "$+$" and "$-$" signs in Eq.~(\ref{abg3}) correspond to the outgoing and ingoing photons in radial direction and in Eq.~(\ref{abg4}) correspond to the photons moving to the north ($\theta=0$) and south ($\theta=\pi$) poles, respectively. }
The above geodesic equations indicate the propagation of light in the rotating ABG spacetime. To determine the unstable circular orbits we introduce $\xi=L_{z}/E$ and $\eta=\mathcal{K}/E^2$. The condition for the unstable circular orbits is given by $\mathcal{R}(r)=0$ and $d\mathcal{R}(r)/dr=0$. Hence, from Eq.~(\ref{R})
\begin{eqnarray}\label{Rz}
\left(r^2+a^2-a \xi\right)^2-\left[\eta +(\xi-a)^2\right] \left(r^2 f(r)+a^2\right)=0, \nonumber \\
\end{eqnarray}
\begin{eqnarray}\label{dRz}
-\left[\eta +(\xi-a)^2\right] \left(2r f(r)+r^2 f'(r)\right) \nonumber \\
+4 r \left(r^2+a^2-a \xi \right)&=&0.
\end{eqnarray}
Now we can easily obtain the expressions for the parameters $\xi$ and $\eta$ from Eq.~(\ref{Rz}) and (\ref{dRz}). These parameters takes the following simple form
\begin{equation}
\xi =\frac{(r^2+a^2)(r f'(r)+2 f(r))-4 (r^2 f(r)+a^2)}{a \left(r f'(r)+2f(r)\right)},
\end{equation}
\begin{equation}
\eta=\frac{r^3 \left[8 a^2 f'(r)-r \left(r f'(r)-2 f(r)\right)^2\right]}{a^2 \left(r f'(r)+2 f(r)\right)^2},
\end{equation}
where $r$ is the radius of the unstable circular orbits and
\begin{eqnarray}
f'(r)&=&-\frac{4 Q^2 r^3}{(Q^2+r^2)^3}+\frac{6 M r^3}{(Q^2+r^2)^{5/2}}+\frac{2 Q^2 r}{(Q^2+r^2)^2}\nonumber \\ &-&\frac{4 M r}{(Q^2+r^2)^{3/2}}.
\end{eqnarray}
These two equations determine the contour of the shadow in the ($\xi$,$\eta$) plane. Furthermore, the parameters $\xi$ and $\eta$ satisfy the following relation
\begin{eqnarray}
\xi^2 + \eta &=& 2 r_{0}^2+a^2+\frac{16 \left(r_{0}^2 f(r_{0})+a^2\right)}{\left(r_{0} f'(r_{0})+2 f(r_{0})\right)^2}\nonumber \\ &-&\frac{8 \left(r_{0}^2 f(r_{0})+a^2\right)}{r_{0} f'(r_{0})+2 f(r_{0})}.
\end{eqnarray}
If we assume that $a=0$ and $Q=0$ then it corresponds to the
Schwarzschild black hole, and the above relation reduces to
\begin{eqnarray}
\xi^2 + \eta &=& \frac{2 r_{0}^{2} (r_{0}^2-3)}{(r_{0}-1)^2}.
\end{eqnarray}
The shape of the critical curve for Schwarzschild black hole is well
known since for this case we have $r_{0}=3$, therefore
$\eta=27-\xi^2$.
\begin{figure}
 \includegraphics[width=1.1\linewidth]{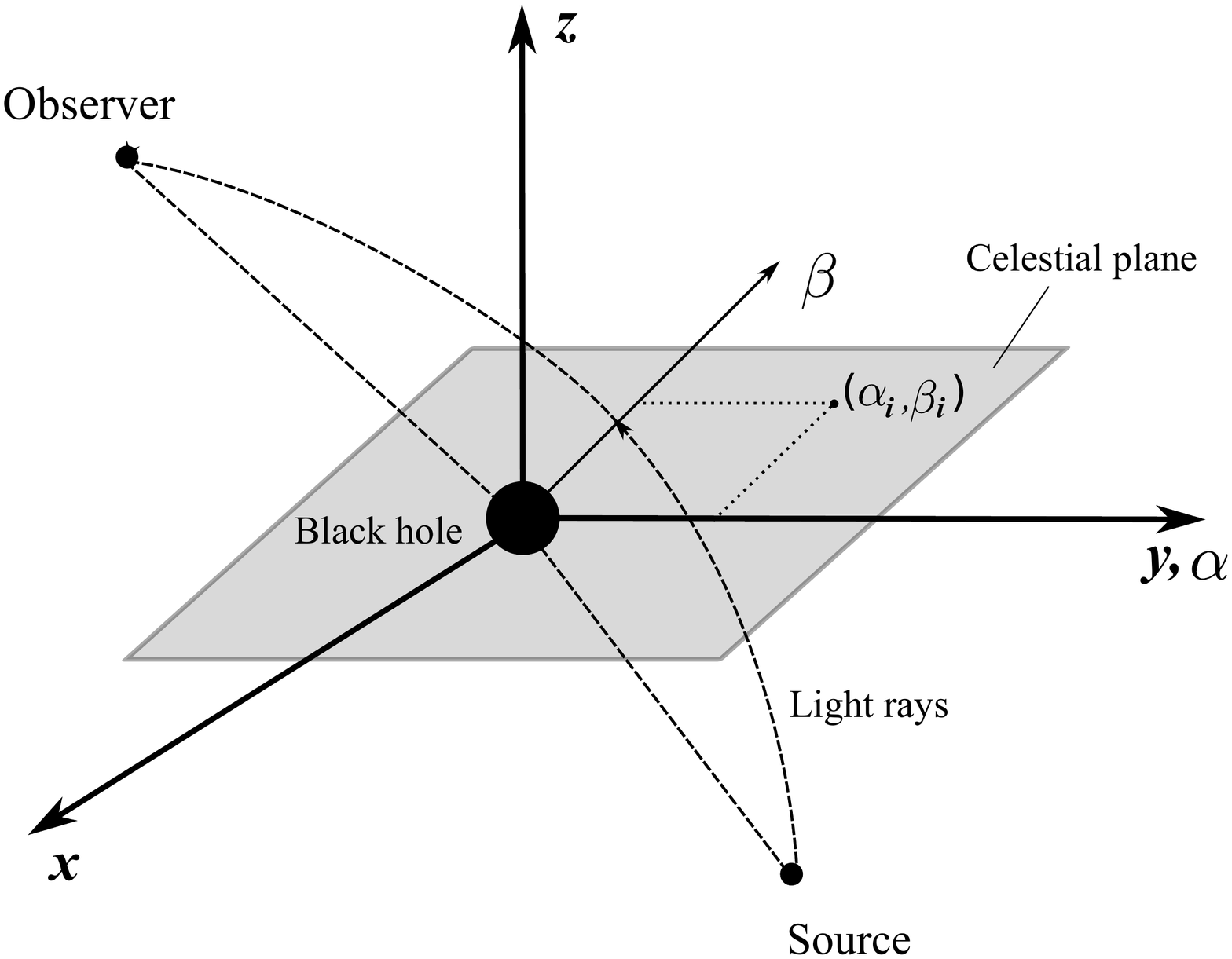}

    \caption{\label{fig35} Schematic illustration of the distant observer's celestial plane and celestial coordinates. }
\end{figure}
\begin{figure*}
    \includegraphics[width=0.245\linewidth]{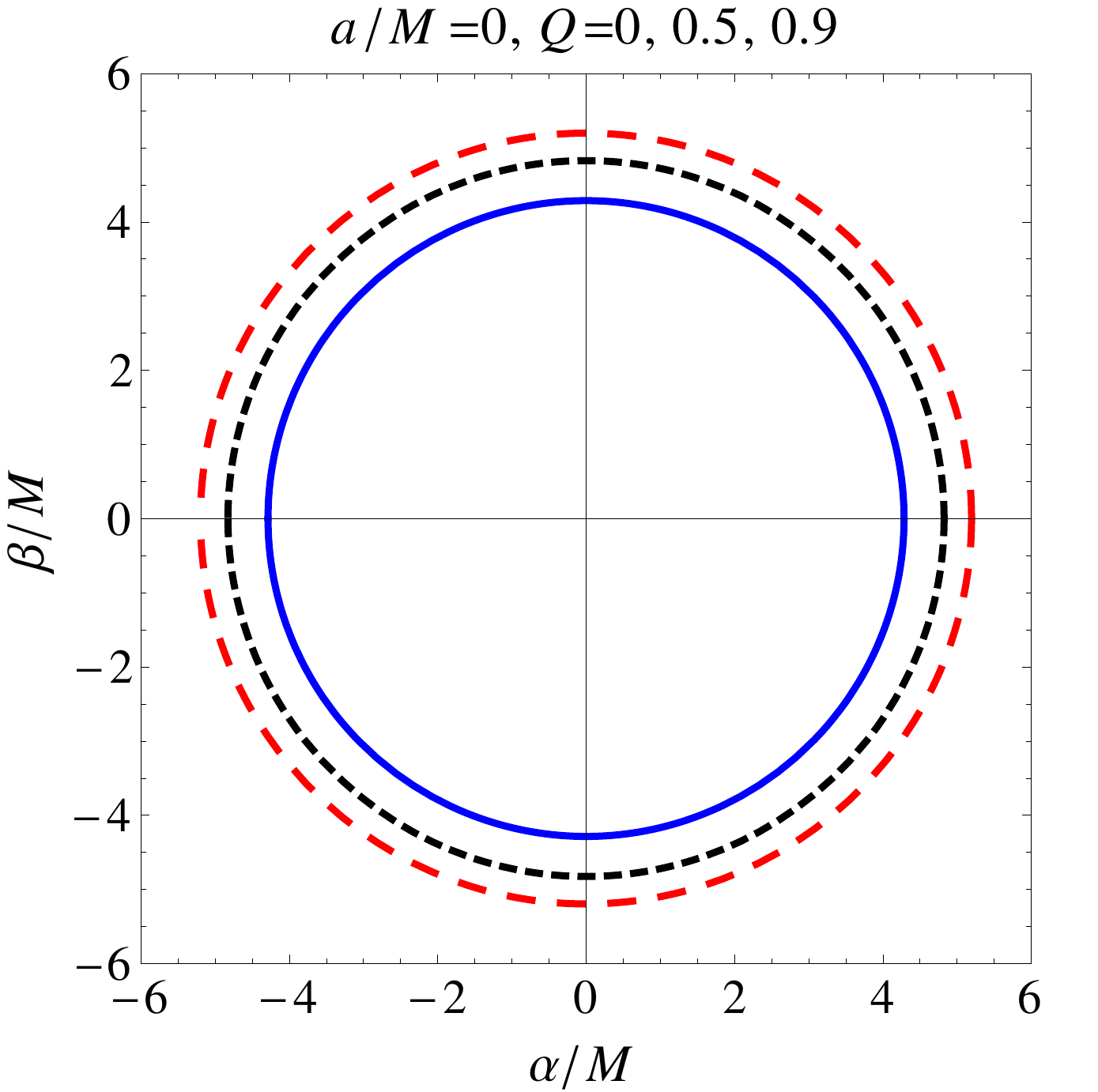}
    \includegraphics[width=0.245\linewidth]{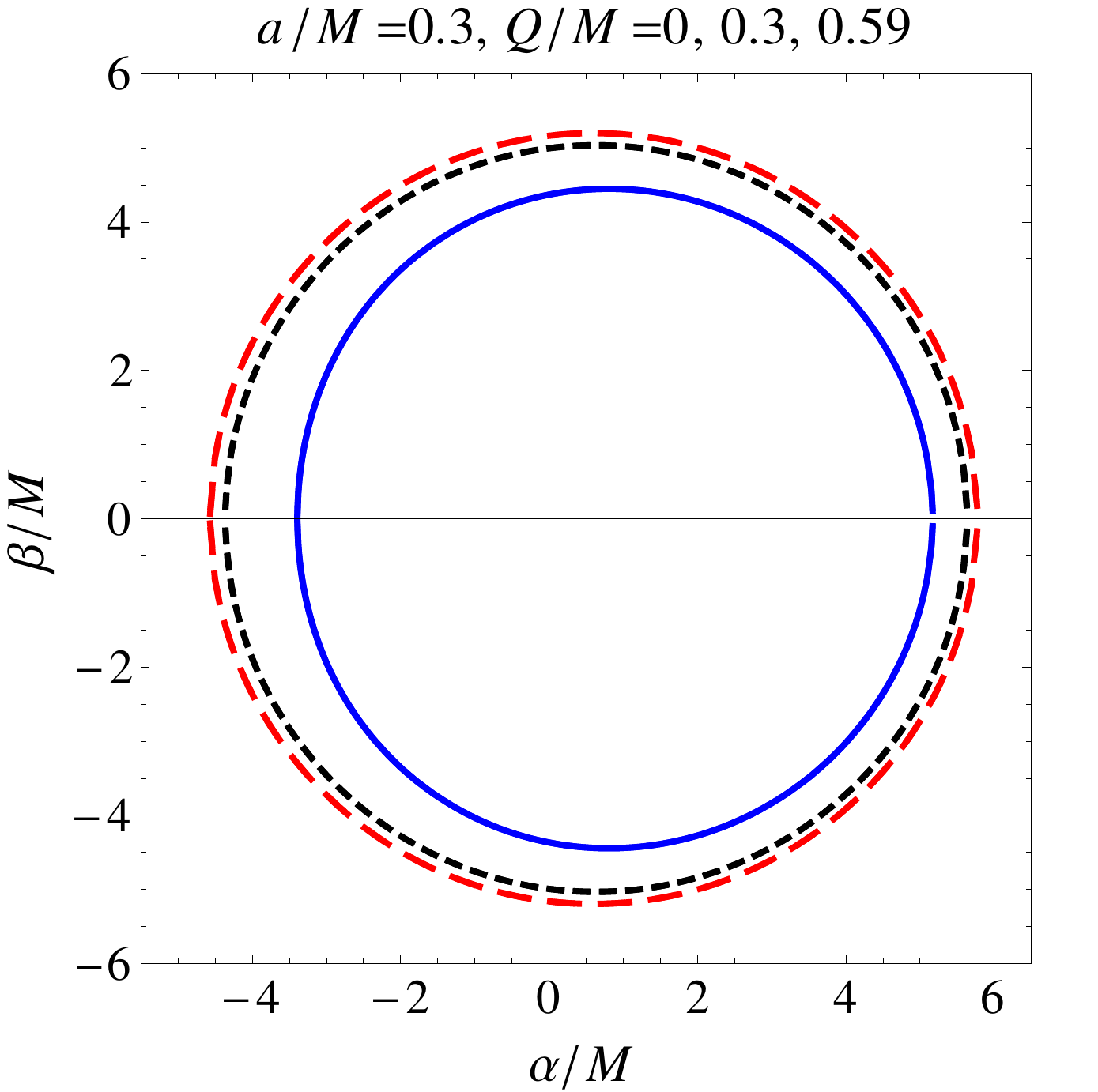}
    \includegraphics[width=0.245\linewidth]{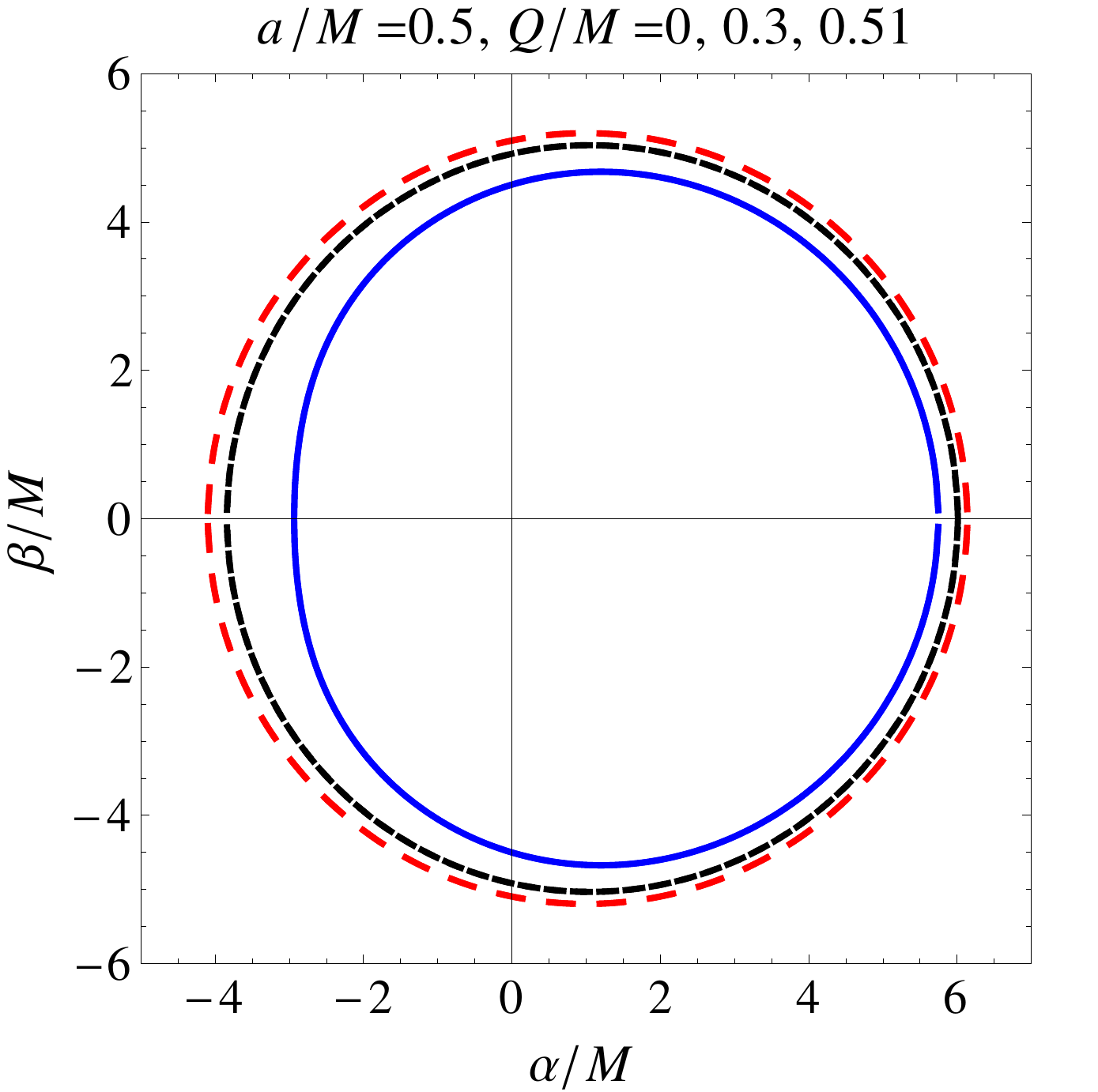}
    \includegraphics[width=0.245\linewidth]{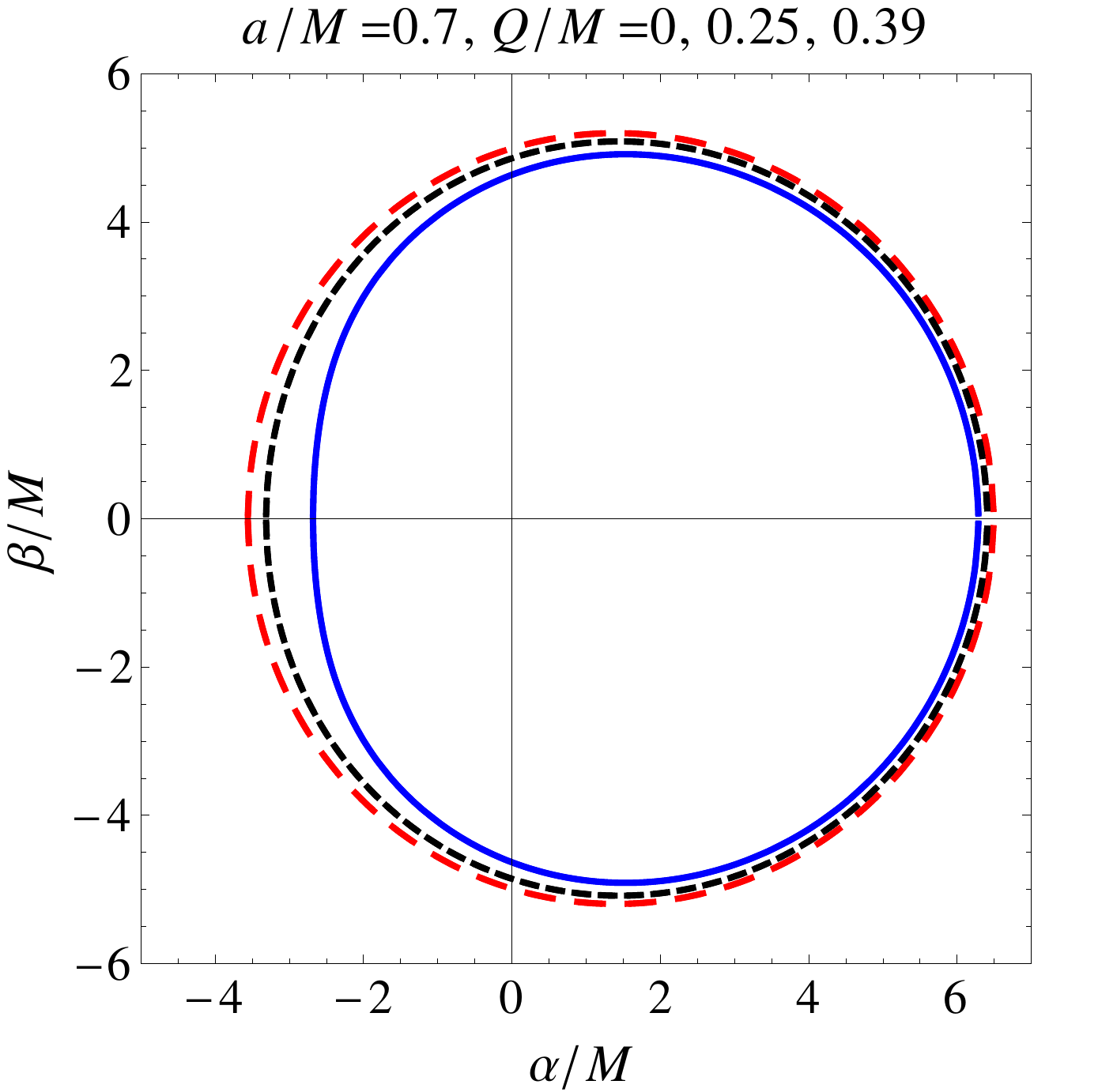}

    \caption{\label{fig1} Plot showing the silhouette of the shadow cast by the rotating ABG black hole for different values of parameters $a/M$ and $Q$. The top left panel corresponds to the static black hole where radius of black hole decrease with increase of parameter~$Q$. }
\end{figure*}

\subsection{Shadow of the rotating ABG black hole}

Now we plan to determine the apparent shape of the rotating ABG black hole shadow. We consider celestial coordinates $\alpha$ and $\beta$ to find the location of the shadow for a better visualization. The coordinate $\alpha$ corresponds to the apparent perpendicular distance of the shape as seen from the axis of symmetry, and the coordinate $\beta$ is the apparent perpendicular distance of the shape from its projection on the equatorial plane. {The schematic illustration of the celestial coordinates is presented in Fig.~\ref{fig35}.} The apparent shape of the black hole shadow for an observer which is far away from the black hole can be given {by the celestial coordinates $\alpha$ and $\beta$}
\begin{eqnarray}
\alpha&=& \lim_{r_{0} \rightarrow \infty}\left(-r_{0}^2 \sin \theta_{0} \frac{d\phi}{dr}\right),
\label{defalpha}\\
\beta&=&\lim_{r_{0} \rightarrow \infty}\left(r_{0}^2\frac{d\epsilon}{dr}\right), \label{defbeta}
\end{eqnarray}
where {according to the standard procedure $r_0$ is the distance
from the black hole to far observer, the celestial coordinates $\alpha$ and $\beta$ are responsible for the
apparent perpendicular distance between the bright image around the black hole due to the light rays fall into the event horizon and i) the
symmetry axis, ii) its projection
on the equatorial plane, respectively, }
$\theta_{0}$ is the angle between the rotation axis of the black hole and the line of sight of
the observer. Furthermore, we calculate $d\phi/dr$ and $d\epsilon/dr$, and substitute these values in the above expressions for the limit $r \rightarrow \infty$, then we have
\begin{equation}
\alpha=-\xi,
\end{equation}
and
\begin{equation}
\beta=\pm \sqrt{\eta}.
\end{equation}
The silhouette of the shadow cast by the rotating ABG black hole can
be visualized from the Fig.~\ref{fig1}. We can see from the
Fig.~\ref{fig1} that for nonrotating case ($a/M=0$), the silhouette of shadow is a perfect
circle and the size of silhouette decreases when the charge $Q$ increases. Furthermore, in a
rotating case ($a/M\neq0$), the silhouette is a deformed circle which
is more deformed if $a/M$ takes an extremal value (cf.~Fig.~\ref{fig1}).

\begin{figure}
 \includegraphics[width=1.1\linewidth]{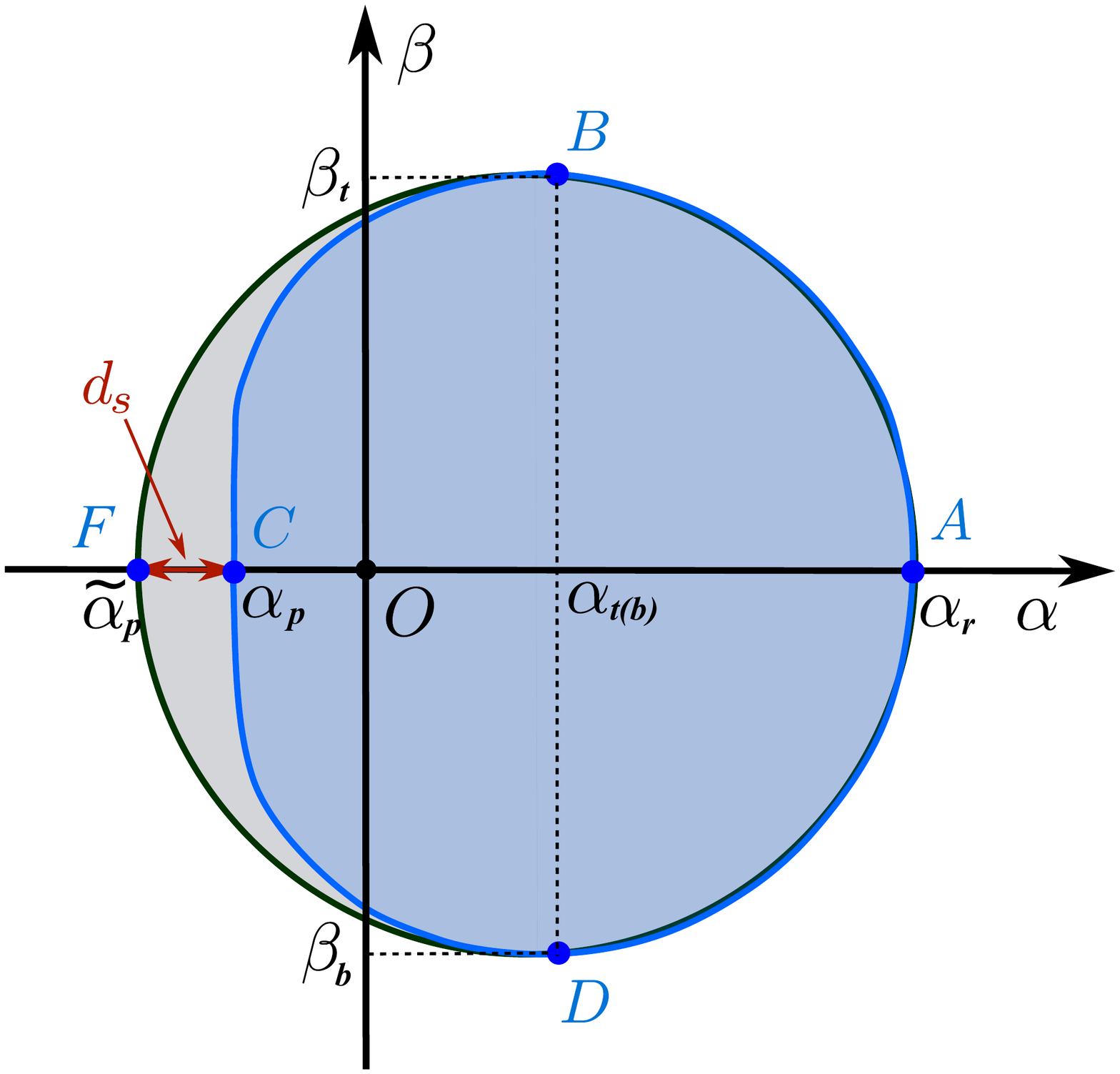}

    \caption{\label{fig45} Schematic representation of the black hole shadow and reference circle. $d_s$ is the distance between the left point of the shadow and the reference circle.}
\end{figure}
\begin{figure*}
    \begin{tabular}{c c}
    \includegraphics[scale=0.58]{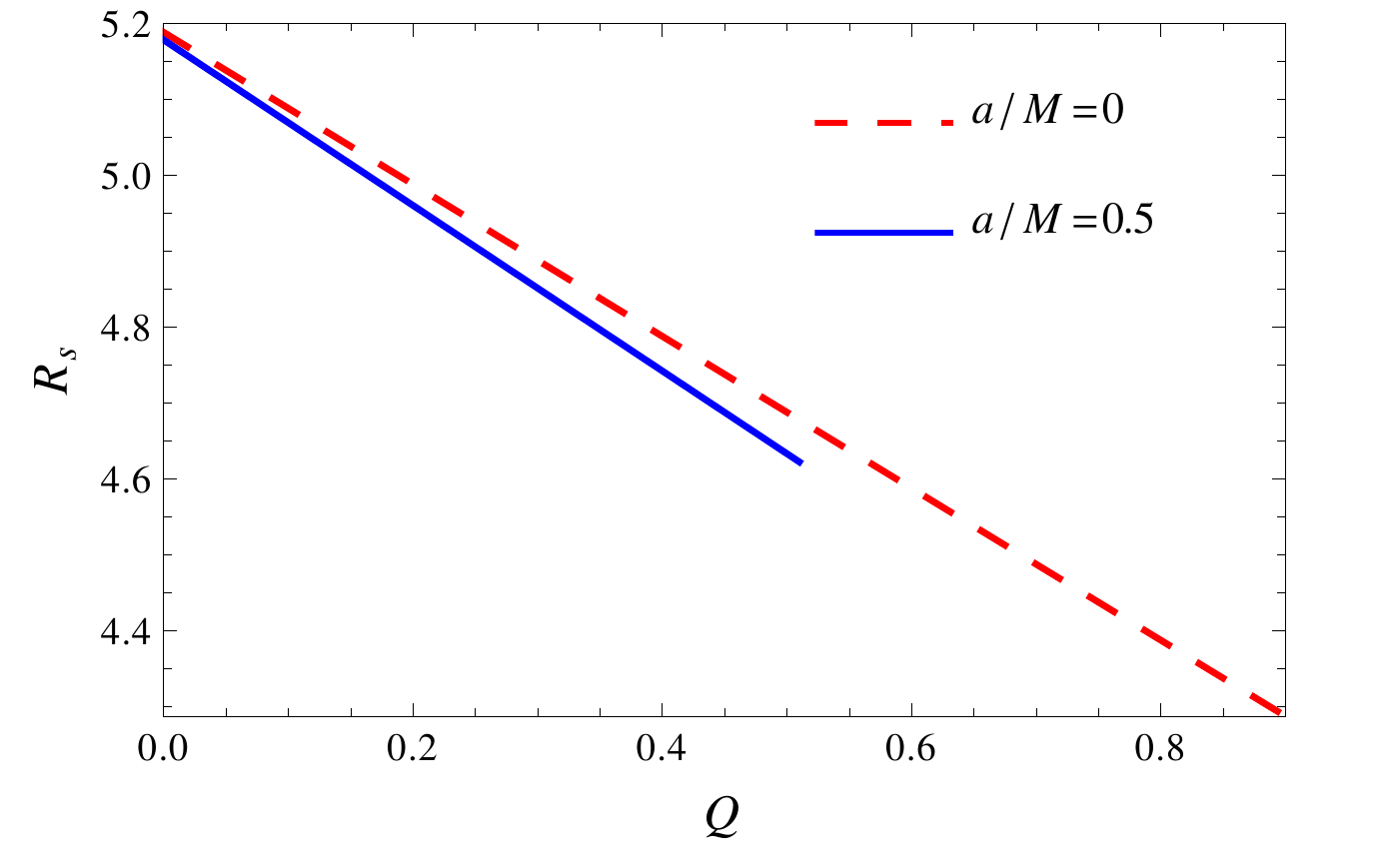}
    \includegraphics[scale=0.58]{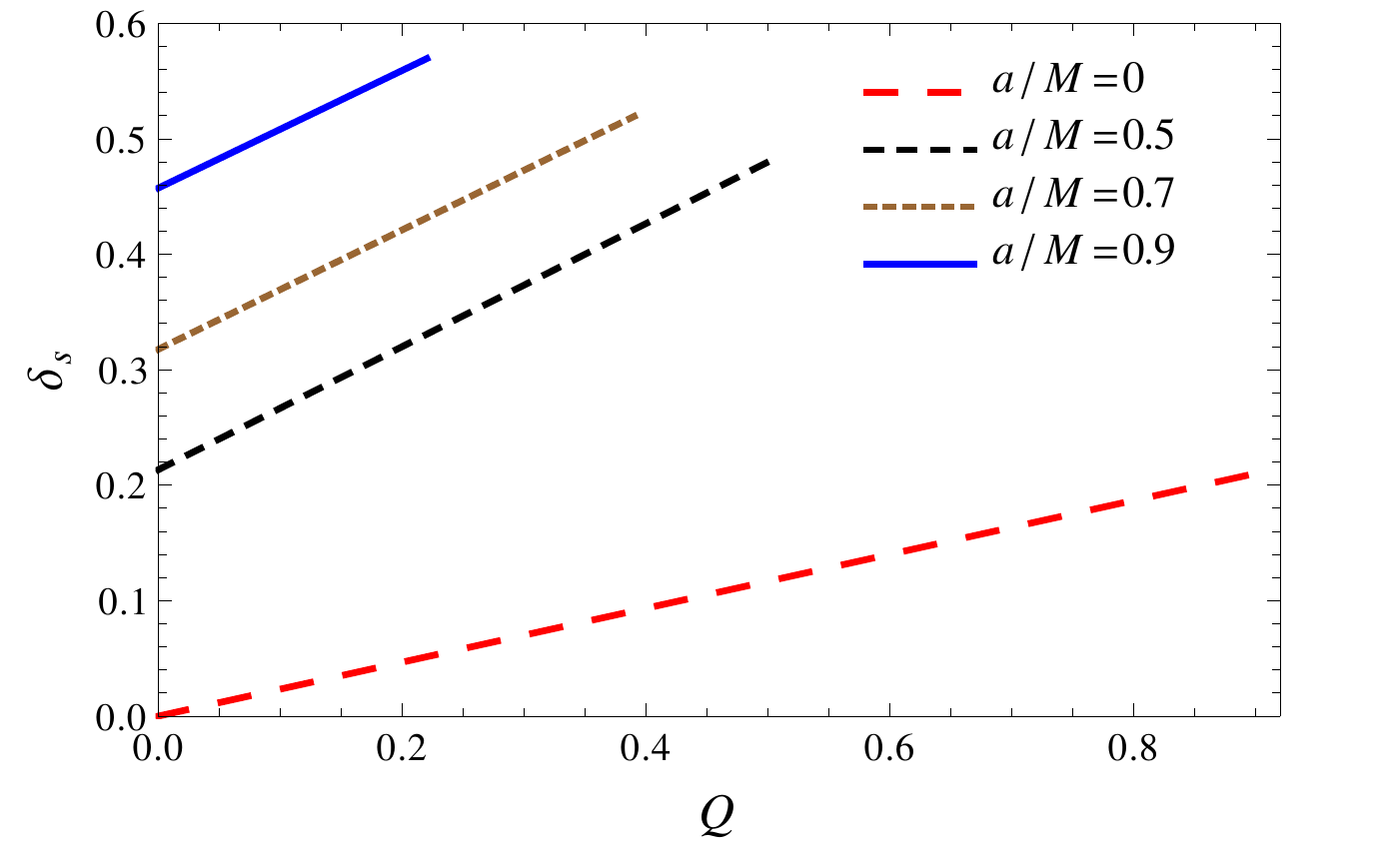}
    \end{tabular}
    \caption{\label{fig2} Plots showing the behavior of $R_{s}$ vs $Q$ (left plot) and $\delta_{s}$ vs $Q$  (right plot) of rotating ABG black hole for the different values of $a/M$.}
\end{figure*}

\begin{figure*}
    \includegraphics[scale=0.6]{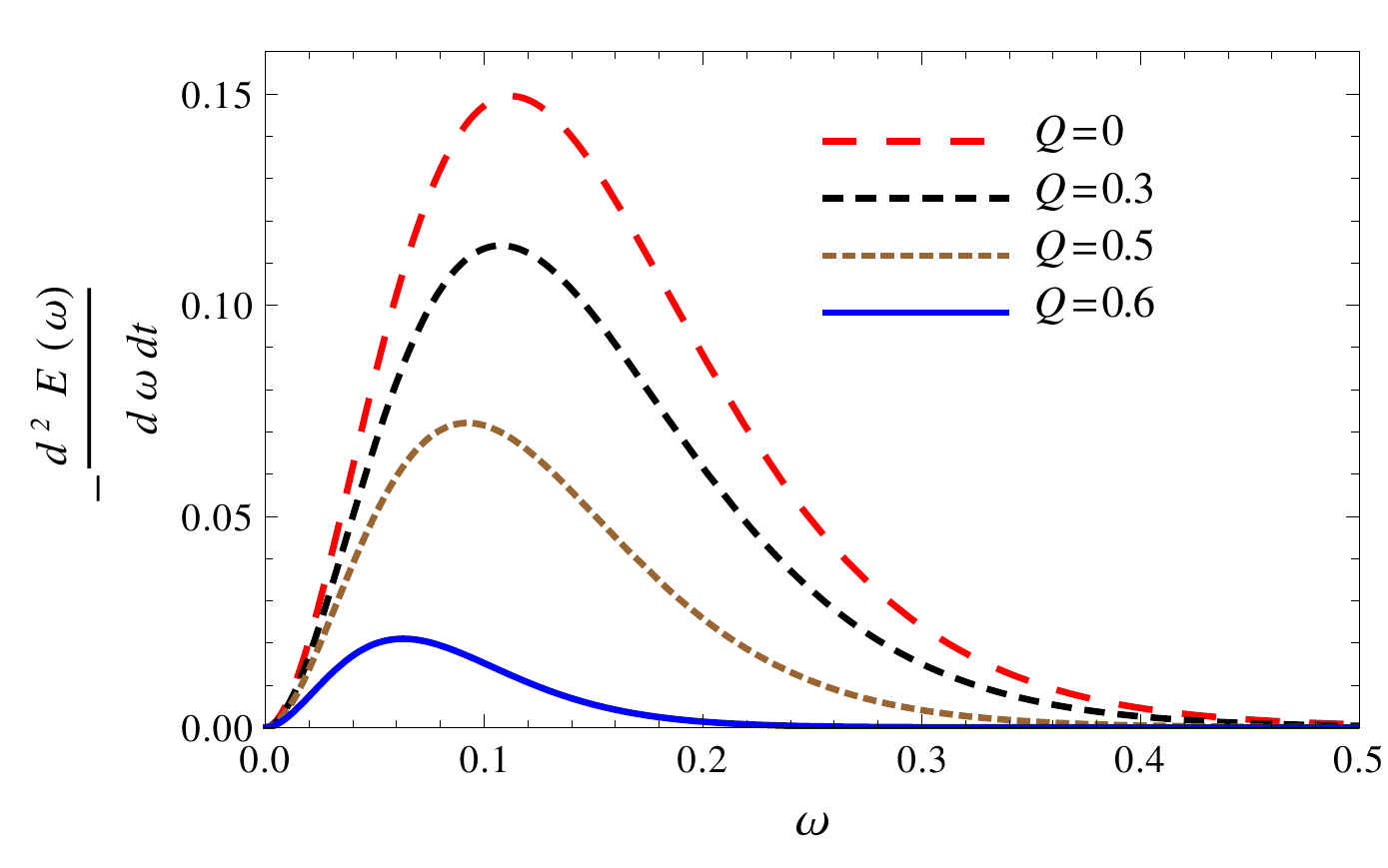}
    \includegraphics[scale=0.6]{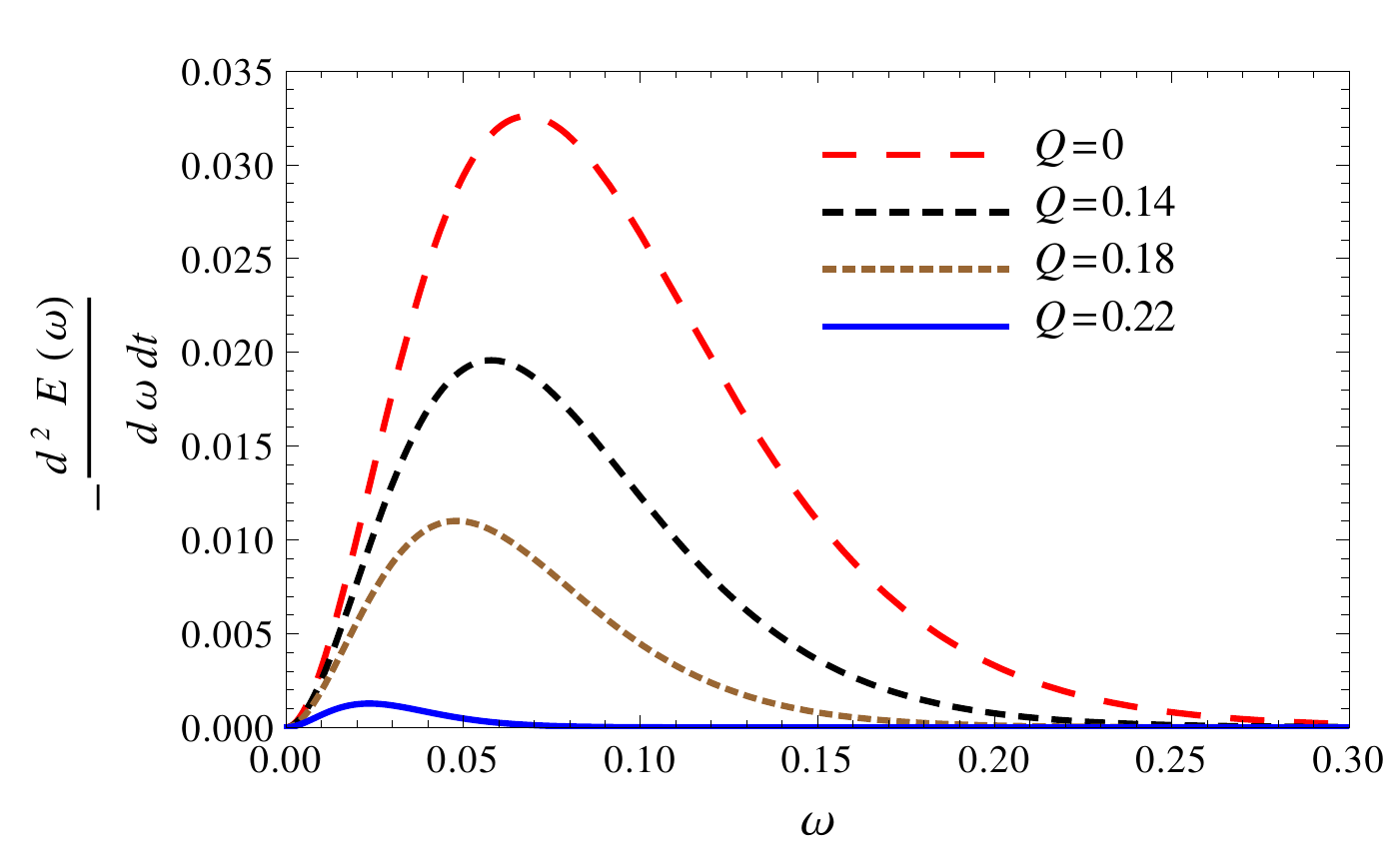}
    \caption{\label{fig3} Plot showing the behavior of energy emission rate versus the frequency ($\omega$). (Left)  For $a/M=0$. (Right) For $a/M=0.9$.}
\end{figure*}

To analyze the shape of the shadow in detail, we define two
astronomical observable: $R_{s}$ which describes the approximate
size of the shadow and $\delta_{s}$ that measures its deformation.
As suggested in Ref.~\cite{Hioki09} the circle of the shadow passing
through the three points B ($\alpha_{t}$,$\beta_{t}$) top one, D
($\alpha_{b}$,$\beta_{b}$) bottom one, and A ($\alpha_{r}$,0) most
right one. {The schematic representation of the above mentioned
definitions is shown in Fig~\ref{fig45}.} The radius or size of the
shadow can be calculated through
\begin{equation}
R_{s}=\frac{(\alpha_{t}-\alpha_{r})^2+\beta_{t}^2}{2\mid \alpha_{t}-\alpha_{r} \mid},
\end{equation}
and the distortion parameter is given as
\begin{equation}
\delta_{s}=\frac{d_s}{R_{s}}=\frac{\tilde{\alpha_{p}}-\alpha_{p}}{R_{s}},
\end{equation}
where the points F ($\tilde{\alpha_{p}}$,0) and C ($\alpha_{p}$,0) cut the horizontal axis at the opposite side of ($\alpha_{r}$,0), $d_s$ is the distance between the left point of the shadow and the reference circle 
(cf. Fig.~\ref{fig45}). 
{We can see the behavior of the observables $R_s$ and $\delta_s$ as a function of charge $Q$ in both nonrotating and rotating black hole cases from the Fig.~\ref{fig2}. It can be observed from the Fig.~\ref{fig2} that the presence of charge $Q$ affects the size of the shadow as well as the distortion parameter, i.e., the size of the shadow decreases and the distortion parameter increases with charge $Q$.
}

\subsection{Energy emission rate of rotating ABG black hole}

{In preceding subsection we have discussed possible visibility of the rotating regular black
through shadow and as we have mentioned in our previous study \cite{Papnoi14} at high energy the  cross
section of the absorption of a black hole slightly modulates near a limiting constant
value. Consequently shadow of the black hole is responsible to high
energy  cross section of absorption by the black hole for the distant observer at far
infinity. The value of the mentioned limiting constant value is derived in terms
of geodesics, and can be analyzed for wave theories. For a
black hole endowed with a photon sphere, the limiting
constant value is the same as the geometrical cross section
of this photon sphere~\cite{Misner73}. Here, the }
limiting constant value of absorption cross section for a spherically symmetric black hole can be given by \cite{Wei13}
\begin{equation}
\sigma_{lim} \approx \pi R_{s}^2,
\end{equation}
and by using this limiting value, we can easily get the energy emission rate in the following form
\begin{equation}
\frac{d^2 E(\omega)}{d\omega dt}= \frac{2 \pi^2 R_{s}^2}{e^{\omega/ T}-1}\omega ^3,
\end{equation}
where $\omega$ represents the frequency of photon and the Hawking temperature ($T$) can be calculated as
\begin{eqnarray}
T &=& -\frac{1}{4 \pi  r_{+} \left(a^2+r_{+}^2\right) \left(Q^2+r_{+}^2\right)^3}[2 Q^6 r_{+}^2+3 Q^4 r_{+}^4 \nonumber \\ &+& Q^2 r_{+}^6-r_{+}^8+a^2 \left(Q^2+r_{+}^2\right)^2 \left(4 Q^2+r_{+}^2\right)],
\end{eqnarray}
{where $r_{+}$ is the outer event horizon of the regular black hole defined as greater root of the solution for the condition  $1/g_{rr}=0$.}
The plots of $d^2 E(\omega)/d\omega dt$ versus $\omega$ can be seen from the Fig.~\ref{fig3} for $a/M=0$ (left panel) and $a/M=0.9$ (right panel).

\section{Rotating Hayward and Bardeen black holes\label{sect3}}

\begin{figure*}

    \includegraphics[width=0.245\linewidth]{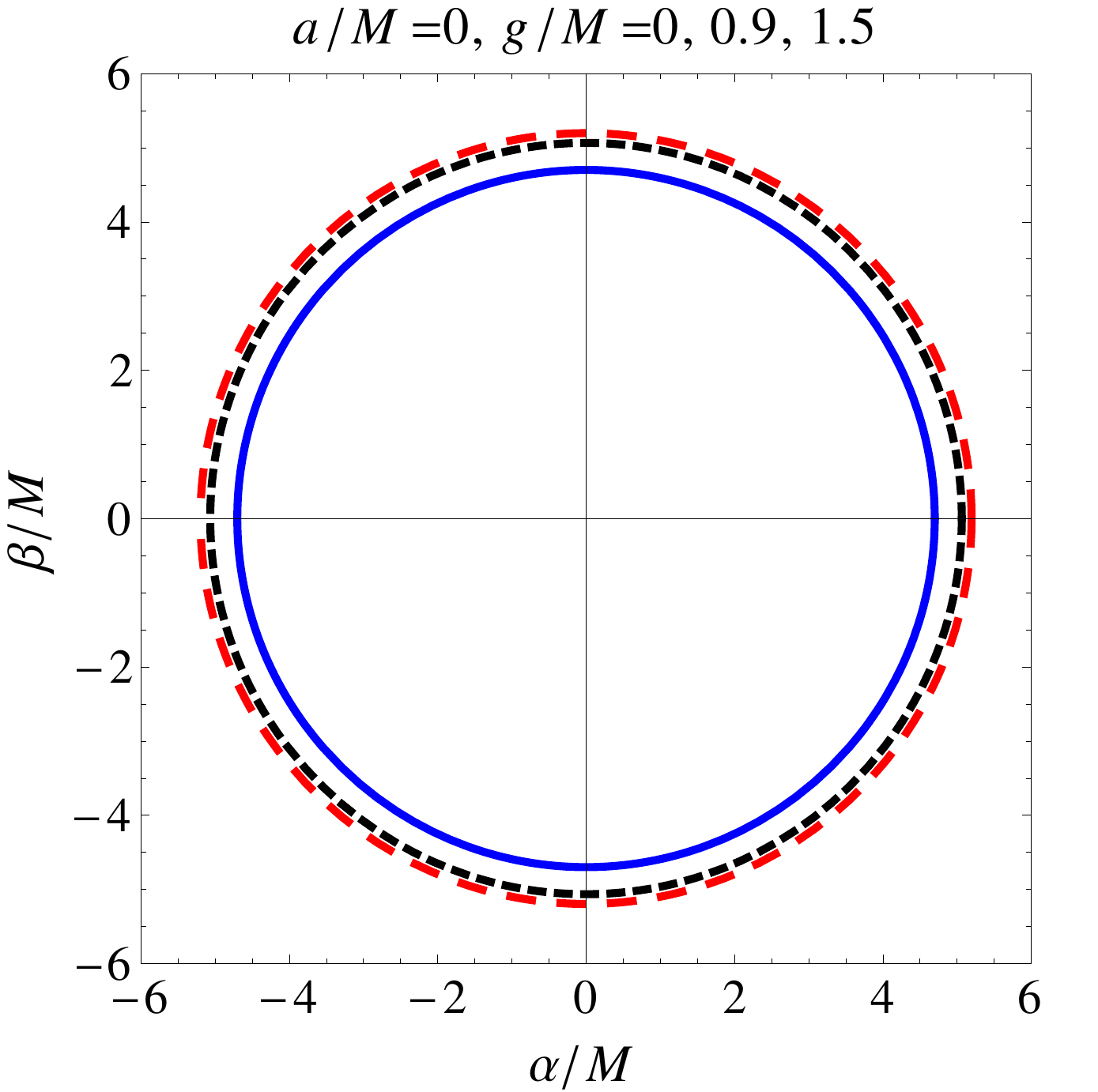}
    \includegraphics[width=0.245\linewidth]{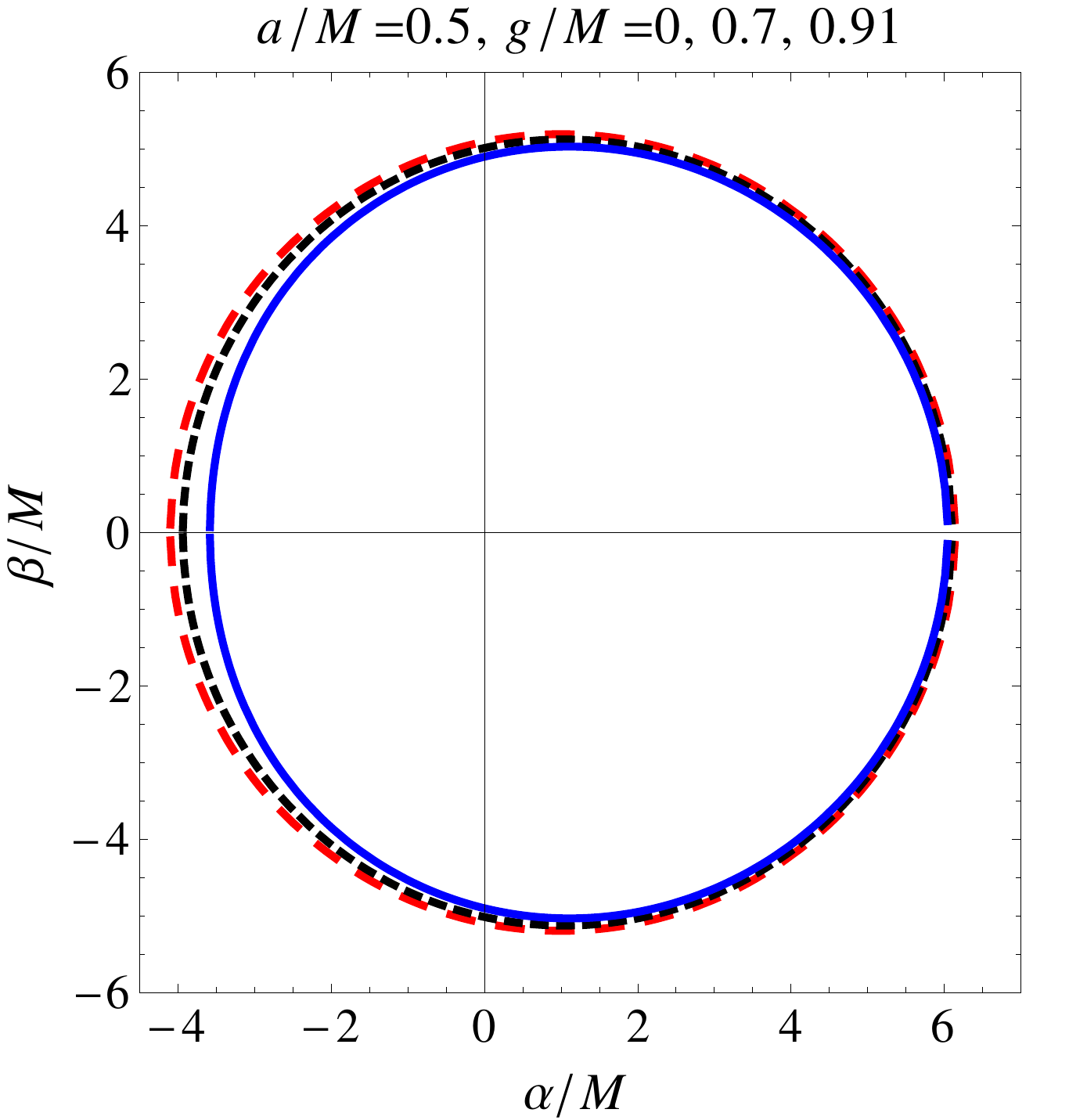}
    \includegraphics[width=0.245\linewidth]{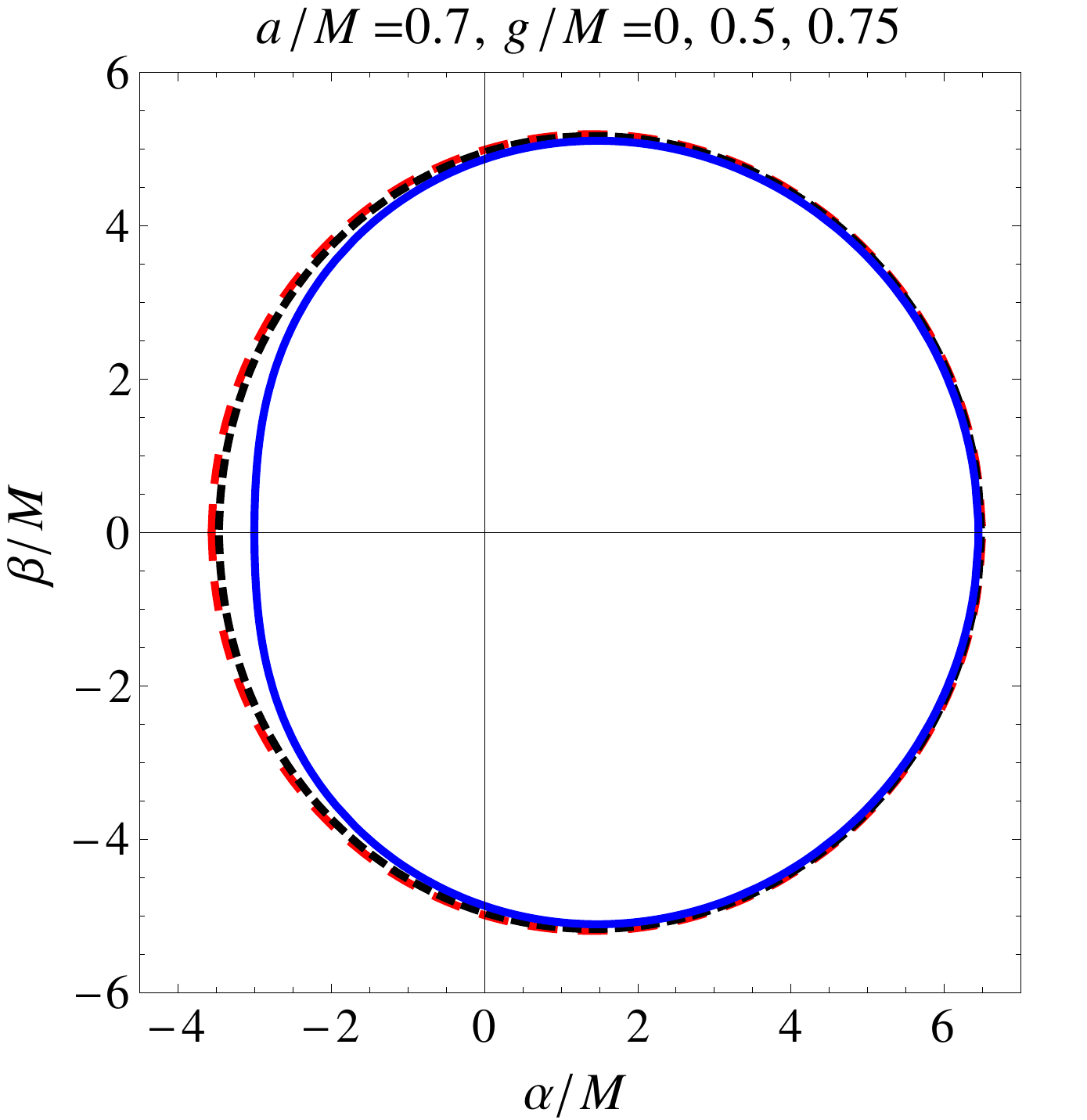}
    \includegraphics[width=0.245\linewidth]{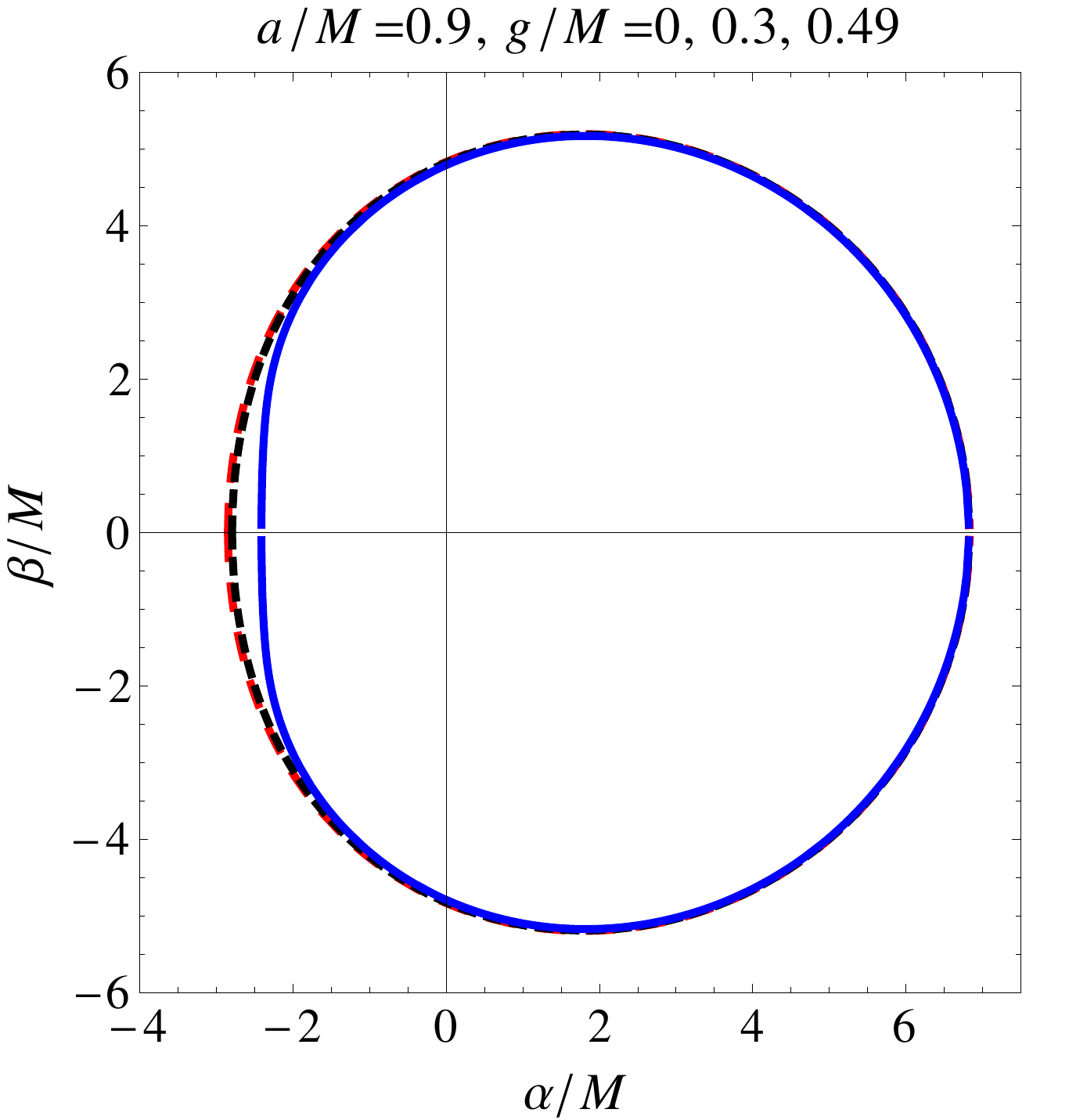}

    \caption{\label{fig4} Plot showing the silhouette of the shadow cast by the rotating Hayward black hole for the different values of the rotation parameter $a/M$. In all plots the outer red lines correspond to $g/M=0$.}
\end{figure*}

\begin{figure*}
    \includegraphics[width=0.245\linewidth]{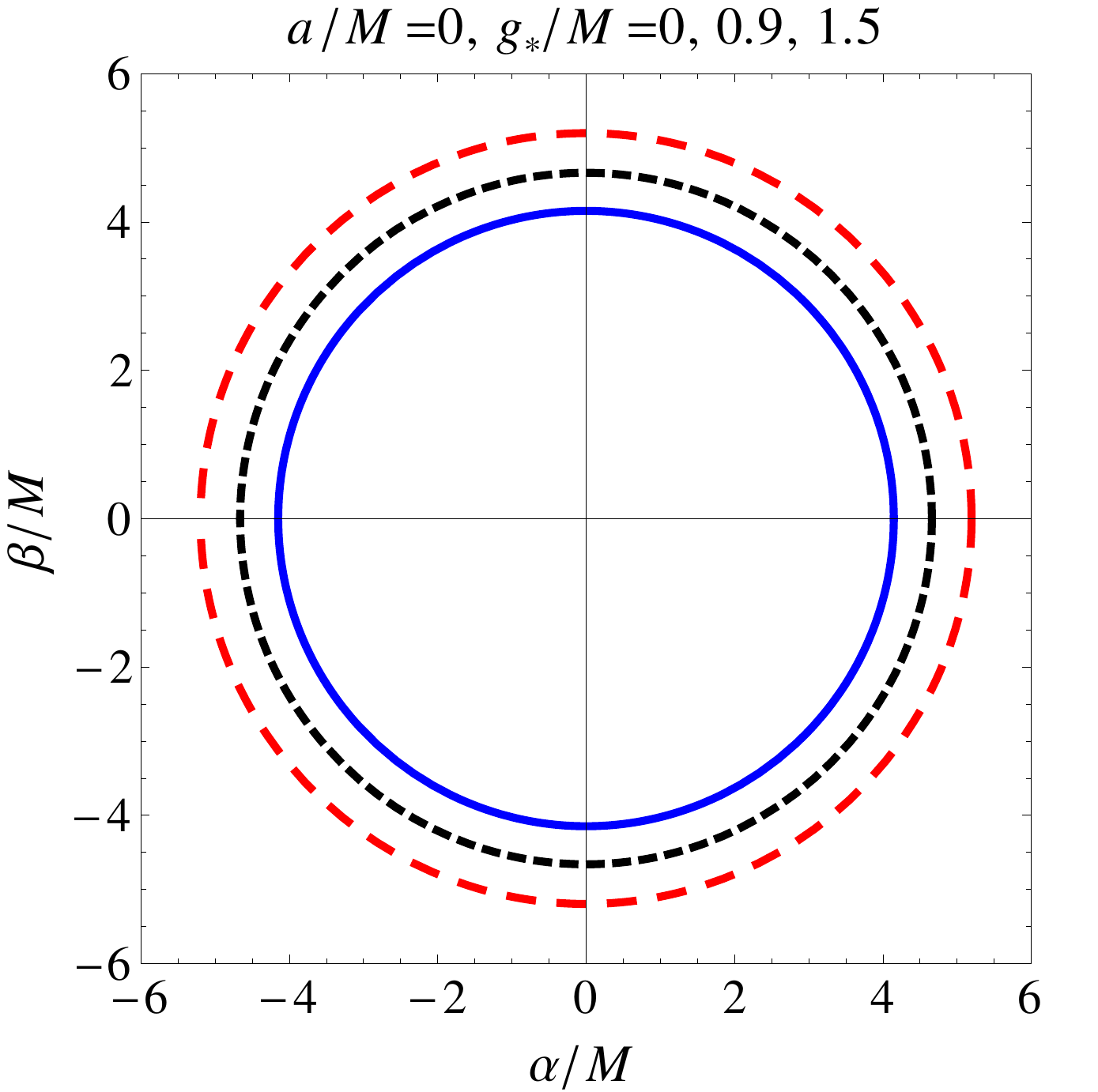}
    \includegraphics[width=0.245\linewidth]{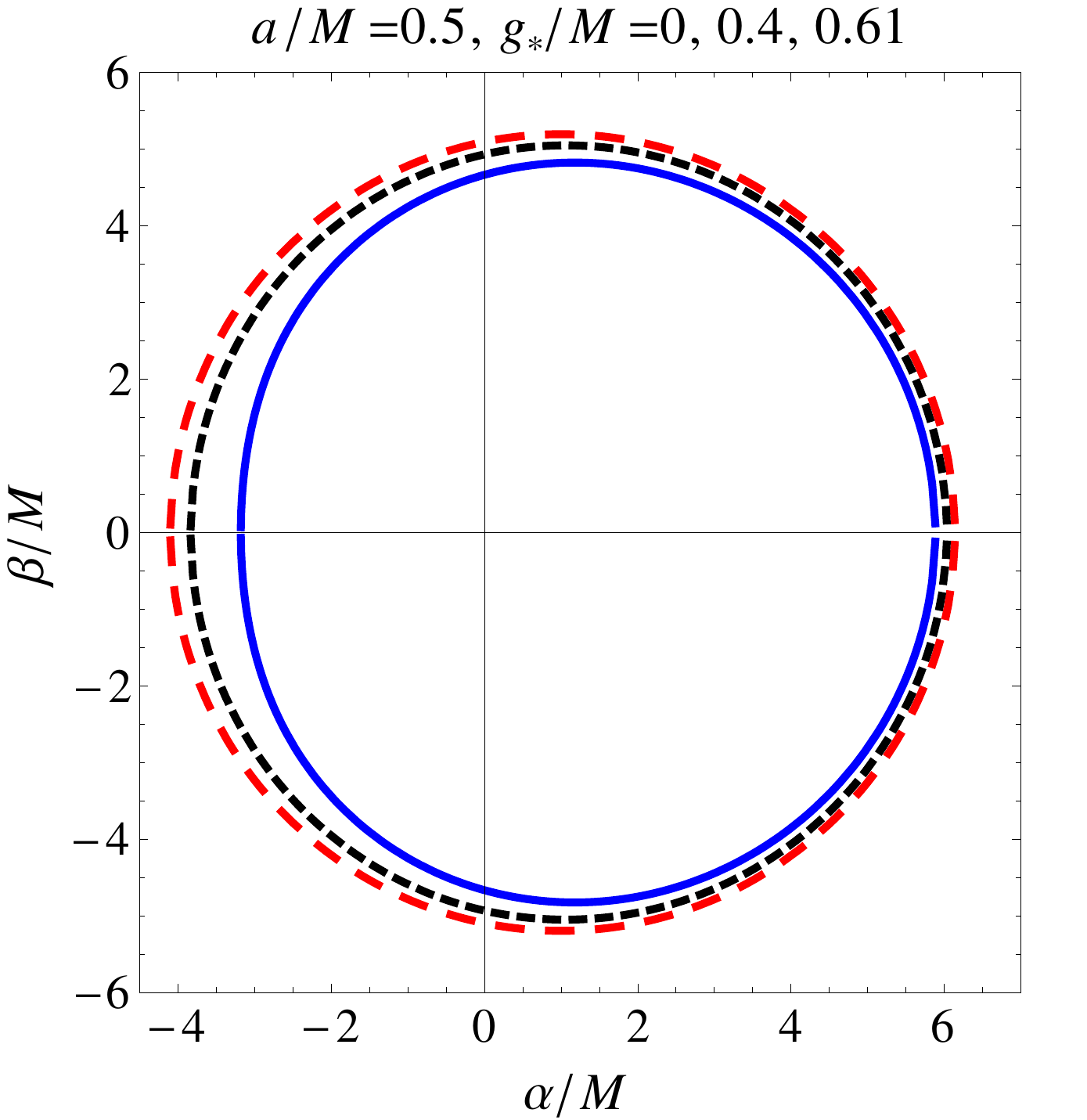}
    \includegraphics[width=0.245\linewidth]{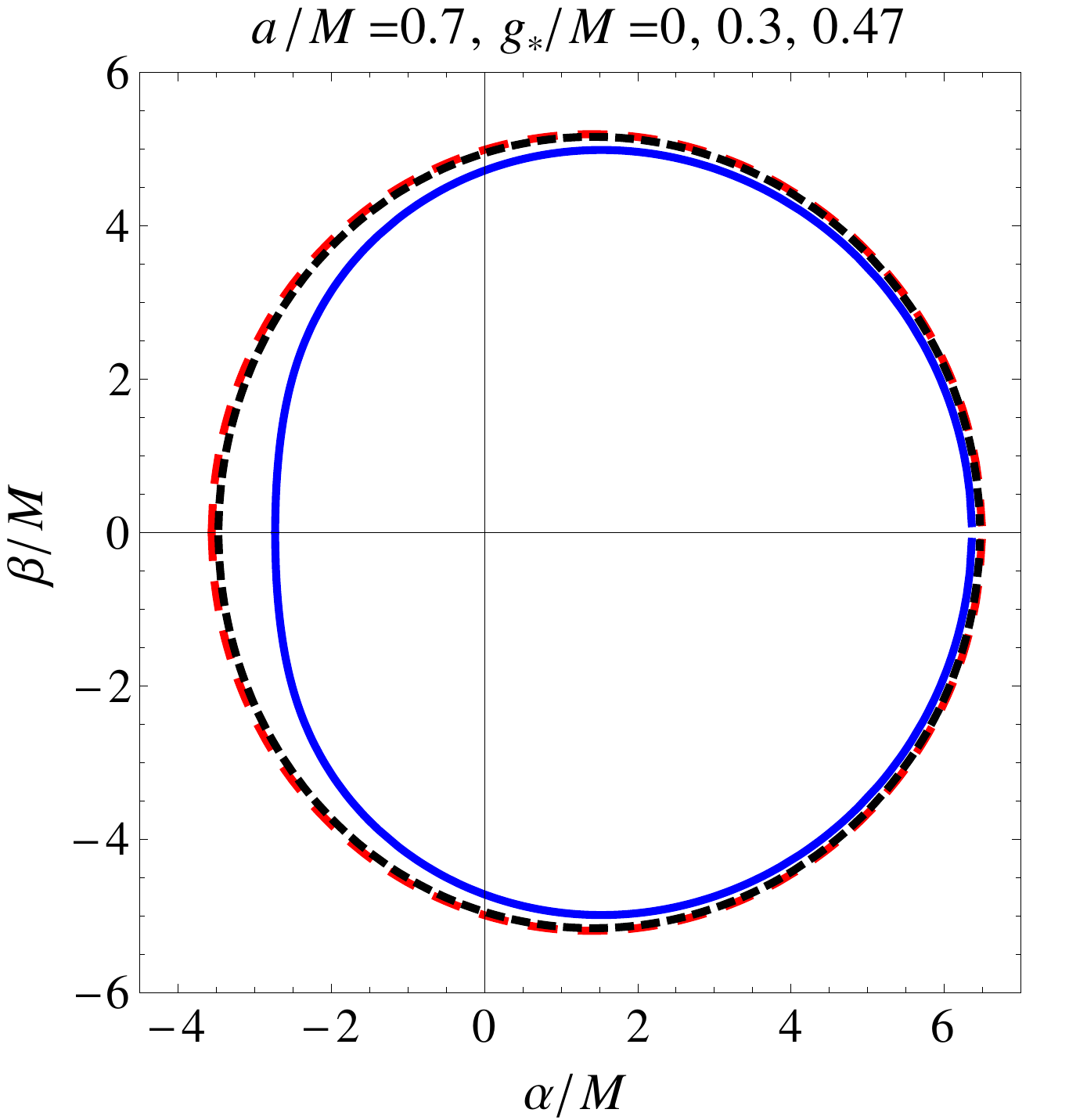}
    \includegraphics[width=0.245\linewidth]{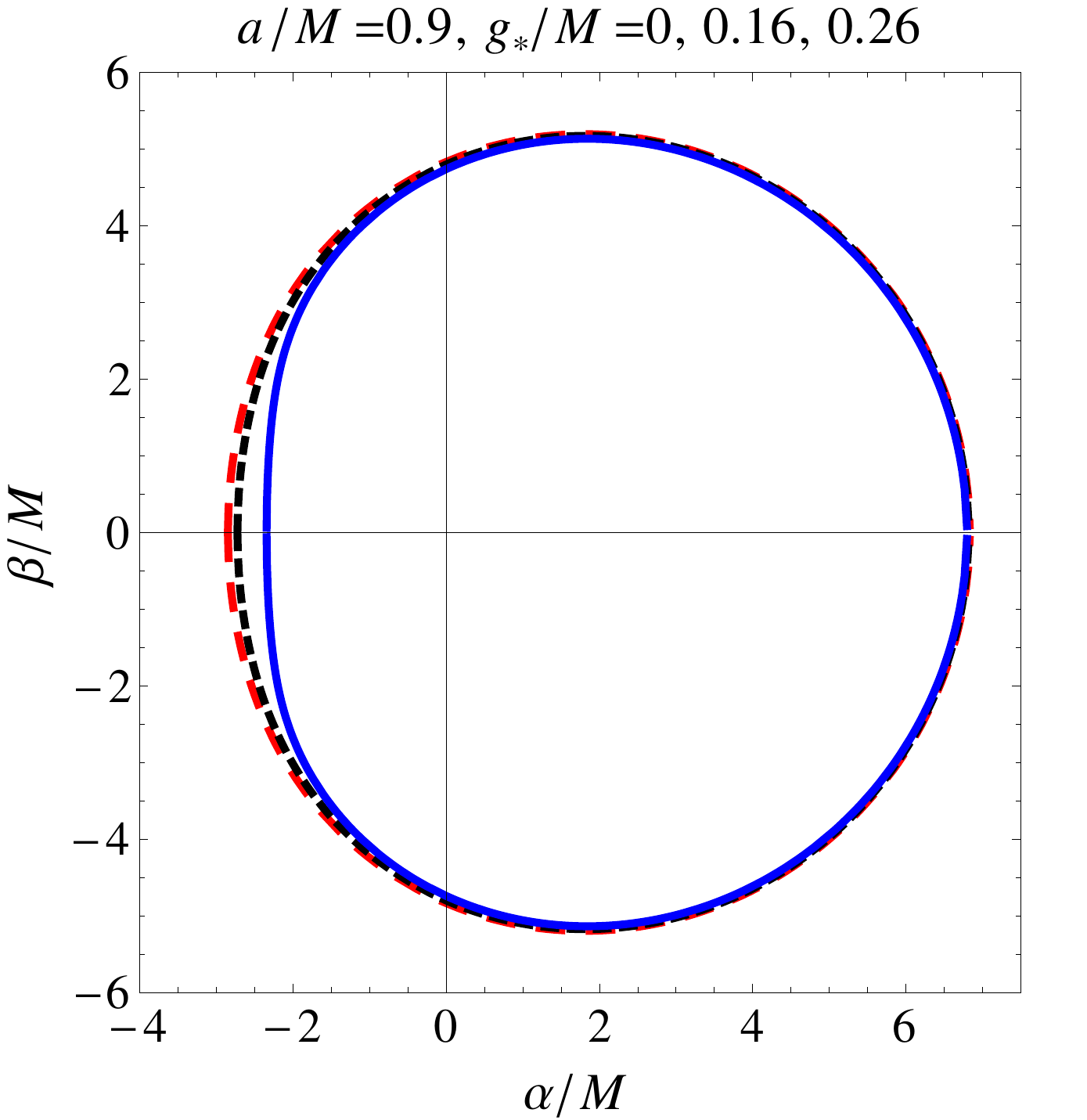}

    \caption{\label{fig6} Plot showing the silhouette of the shadow cast by the Bardeen black hole for the different values of the rotation parameter $a/M$. In all plots the outer red lines correspond to $g_{*}/M=0$}
\end{figure*}

\begin{figure*}
    \includegraphics[scale=0.58]{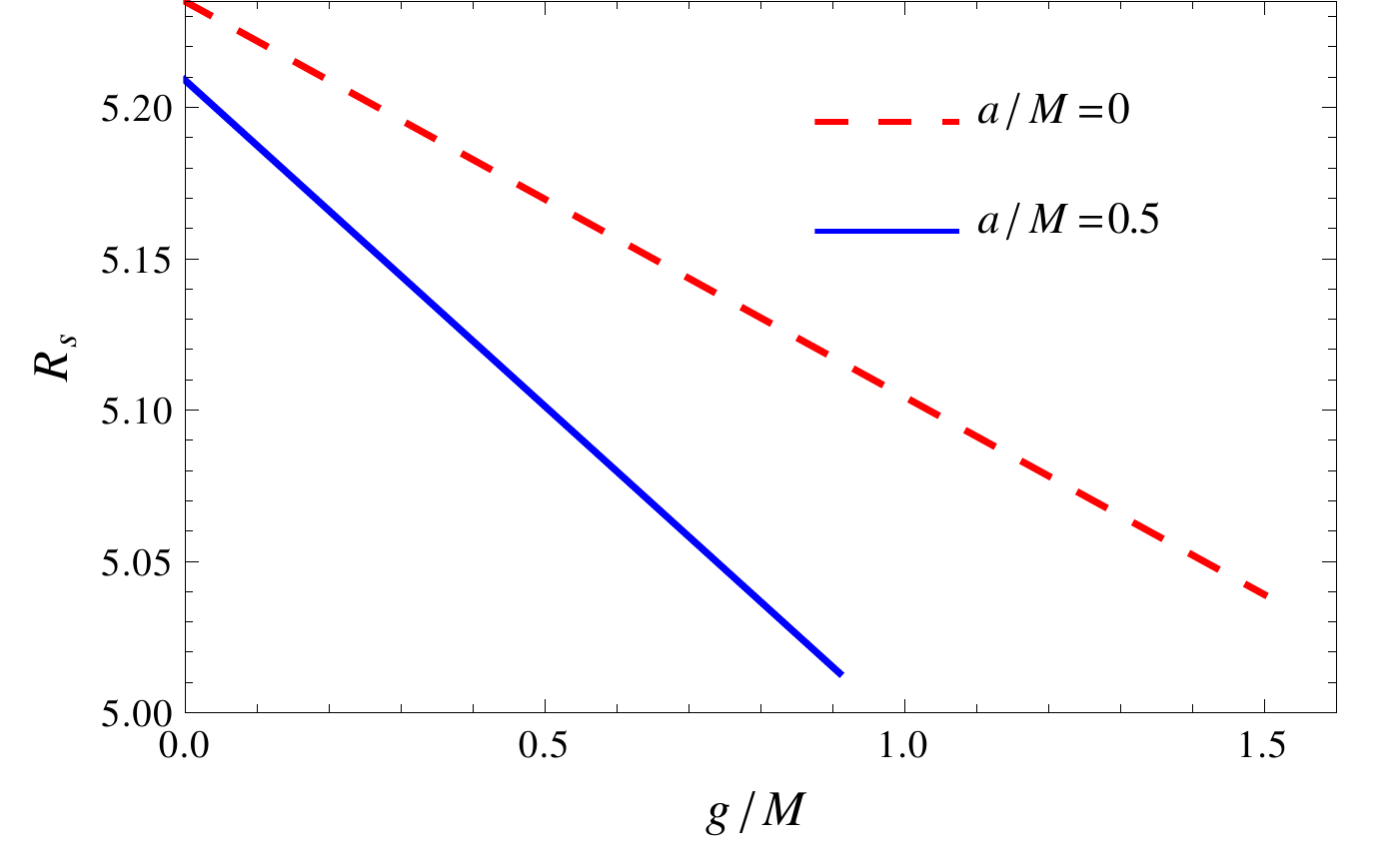}\hspace{0.35cm}
    \includegraphics[scale=0.58]{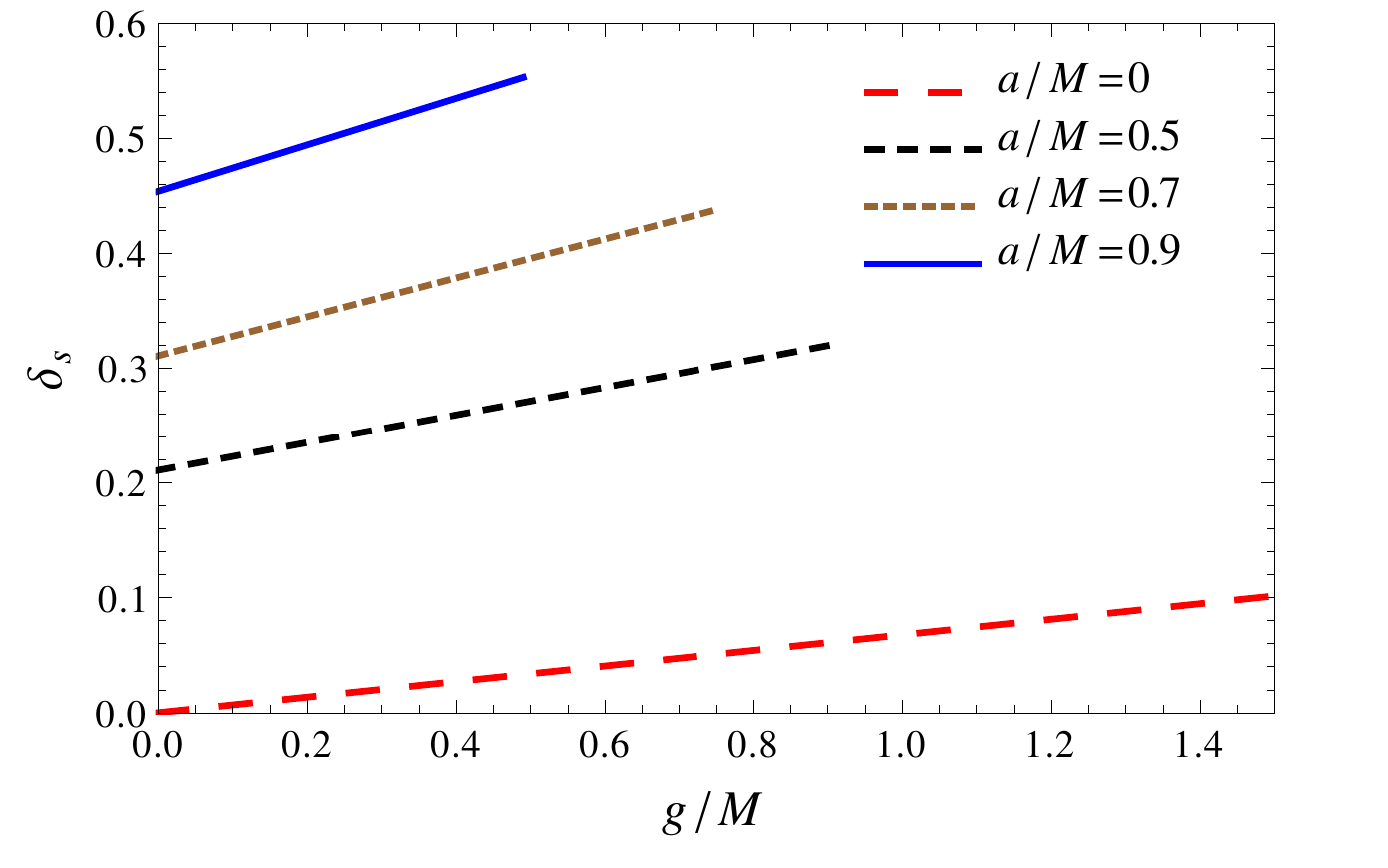}
    \caption{\label{fig5} Plots showing the behavior of $R_{s}$ vs $g/M$ (left panel) and $\delta_{s}$ vs $g/M$ (right panel) of rotating Hayward black hole  for the different values of the rotation parameter $a/M$.}
\end{figure*}

\begin{figure*}
    \includegraphics[scale=0.58]{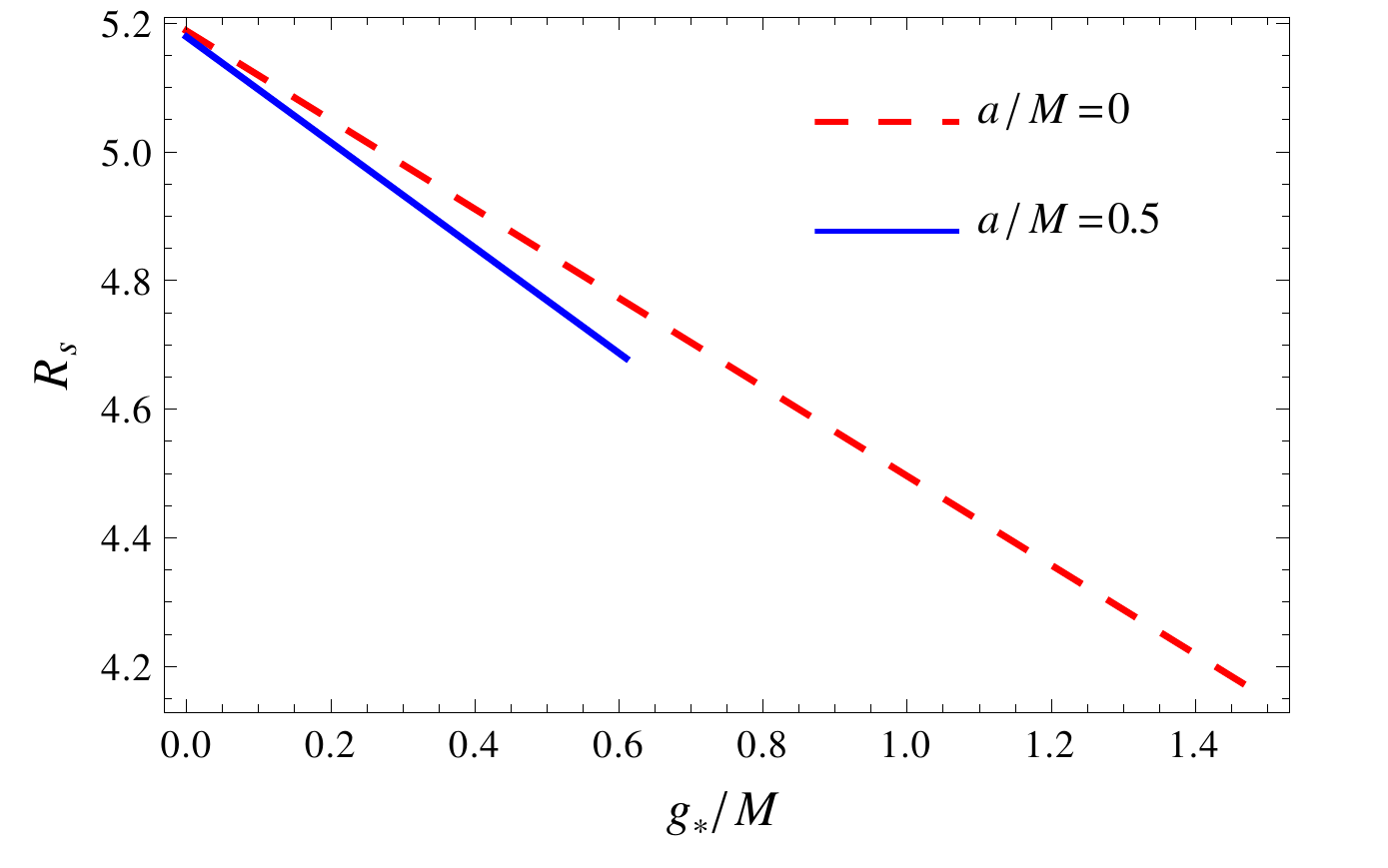}\hspace{0.35cm}
    \includegraphics[scale=0.58]{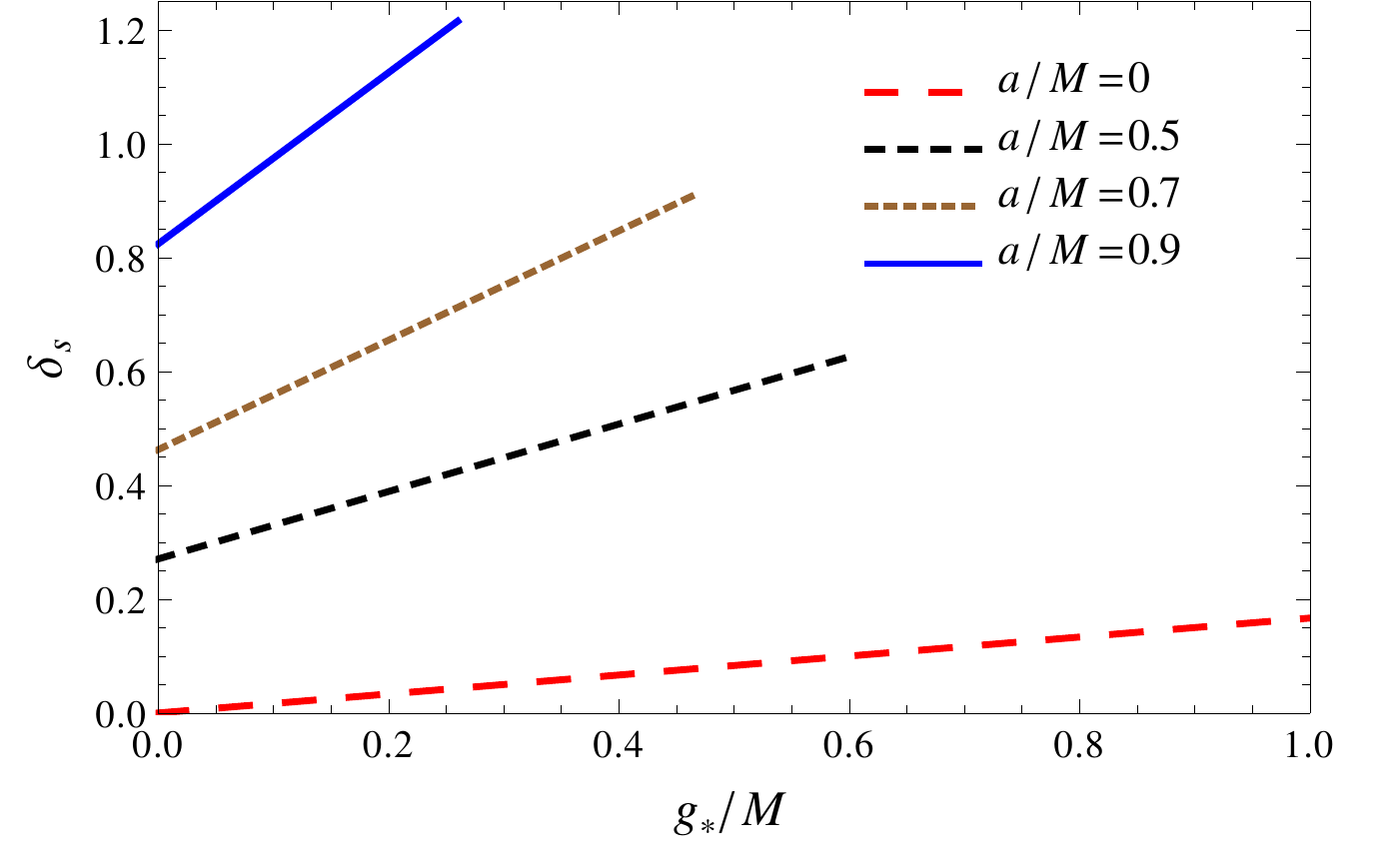}
    \caption{\label{fig7} Plots showing the behavior of $R_{s}$ vs $g_{*}/M$ (left panel) and $\delta_{s}$ vs $g_{*}/M$ (right panel) of rotating Bardeen black hole.}
\end{figure*}

The spacetime metric of the rotating Kerr-like black hole, in the Boyer-Lindquist coordinates is given as
\begin{eqnarray}
\label{metrics}
d{s}^2 &=& -\left(1-\frac{2mr}{\Sigma}\right)dt^2 -\frac{4amr \sin^2 \theta}{\Sigma}dtd\phi +\frac{\Sigma}{\Delta}dr^2
\nonumber \\
&+& \Sigma d\theta^2 +\left(r^2+a^2+\frac{2a^2mr \sin^2 \theta}{\Sigma}\right)\sin^2 \theta d\phi^2,
\nonumber \\
\end{eqnarray}
where
\begin{eqnarray}
\Sigma =r^2 + a^2\cos^2\theta,\;\;\;\;\; \Delta = r^2-2mr+a^2. \label{eqn30}
\end{eqnarray}
The above metric represent Kerr black hole spacetime, if $m \rightarrow M$. In this case, mass $m$ depends on $r$, which is given by
\begin{eqnarray}
\label{mh}
m \rightarrow m_{h} = M\frac{r^3}{r^3+g^3}, \label{eqn31}
\end{eqnarray}
and
\begin{eqnarray}
\label{mb}
m \rightarrow m_{b} = M\left(\frac{r^2}{r^2+g_{*}^2}\right)^{3/2}, \label{eqn32}
\end{eqnarray}
where black hole masses $m_h$ and $m_b$ correspond to the rotating Hayward and rotating Bardeen regular black hole, respectively. The constants $a$, $g$, and $g_{*}$ correspond to the rotation parameter, deviation parameter, and the magnetic charge due to the non-linear electromagnetic field, respectively. The corresponding geodesic equations of these black holes have the same form and the difference is just due to the different mass. One can easily find the equations in the following form
\begin{equation}
\label{eqm1}
\Sigma \frac{d t}{d \sigma} =  a ( L_{z}-aE \sin^2 \theta) +  \frac{r^2 + a^2}{\Delta}\left[(r^2 + a^2)E-aL_{z}\right],
\end{equation}
\begin{equation}
\label{eqm2}
\Sigma \frac{d \phi}{d \sigma} = \left( \frac{L_{z}}{\sin^2 \theta} -aE \right) + \frac{a}{\Delta}\left[(r^2 + a^2)E-aL_{z}\right],
\end{equation}
\begin{equation}
\label{eqm3}
\Sigma \frac{d r}{d \sigma} = \pm \sqrt{\mathcal{R}},
\end{equation}
\begin{equation}
\label{eqm4}
\Sigma \frac{d \theta}{d \sigma} = \pm \sqrt{\Theta},
\end{equation}
where $\sigma$ is an affine parameter, and
\begin{equation}
\label{Rhb}
\mathcal{R}= [(r^2 + a^2)E -aL_{z}]^2-\Delta[\mathcal{K}+(L_{z}-aE)^2],
\end{equation}
\begin{equation}
\label{Thb}
\Theta= \mathcal{K}+\cos^2 \theta \left(a^2E^2-\frac{L_{z}^{2}}{ \sin^2 \theta}\right).
\end{equation}
{Note, that in this Section the equations of motion are not restricted to be at the equatorial plane ($\theta \neq {\rm const}$).} With the help of the condition for the unstable circular orbits of the particles, i.e., $\mathcal{R}(r)=0$ and $d\mathcal{R}(r)/dr=0$, we have
\begin{eqnarray}
\label{R1}
&& r^4+(a^2-\xi^2-\eta)r^2 \nonumber\\&&\quad +2m\left[\eta +(\xi -a)^2\right] r-a^2 \eta =0,\\
&& \label{dR1}
4r^3+2(a^2-\xi^2-\eta)r +2m\left[\eta +(\xi -a)^2\right] \nonumber\\&&\quad -2m'\left[\eta +(\xi -a)^2\right]r =0.
\end{eqnarray}
%%%%%%%%%%%%%%%%%%%%%%%%%%%%%%%%%%%%
By solving the above equations simultaneously, we can easily get the parameters $\xi$ and $\eta$ as
\begin{eqnarray}
\xi&=&\frac{m\left(a^2-3 r^2\right) +r \left(r^2+a^2\right) \left(m'+1\right)}{a \left[m+r \left(m'-1\right)\right]},
\\
\eta &=& -\frac{r^3}{a^2} \bigg[(1+ m'^2)r^3+2 m'(r^2-3 mr +2 a^2)r\nonumber\\ &&-m (6 r^2-9 mr+4a^2)\bigg] \big[m+r (m'-1)\big]^{-2},
\end{eqnarray}
%%%%%%%%%%%%%%%%%%%%%%%%%%%%%%%%%%%%%%
where $m'$ represents the derivative of $m$ with respect to~$r$.

\begin{figure*}
    \includegraphics[scale=0.6]{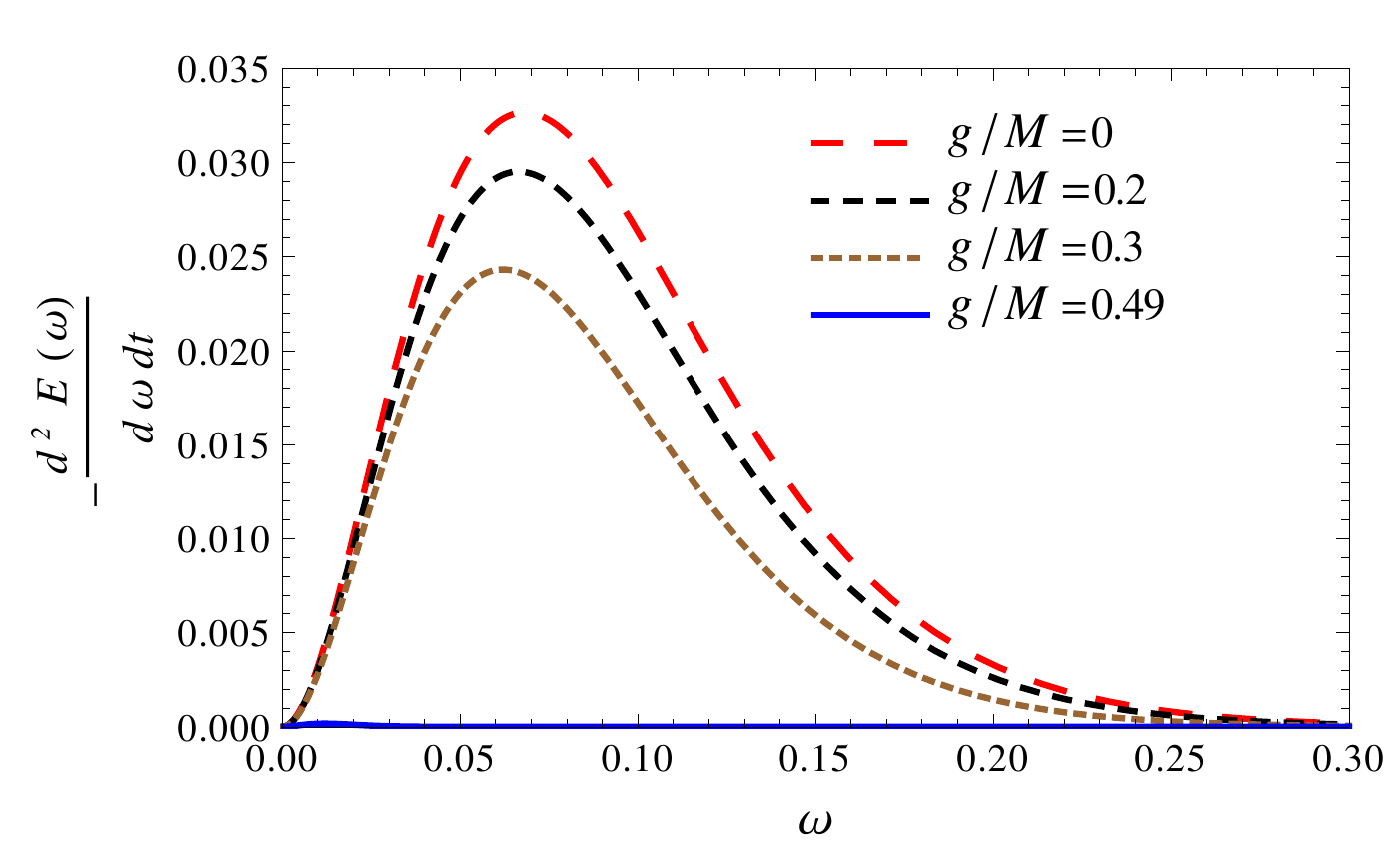}
    \includegraphics[scale=0.6]{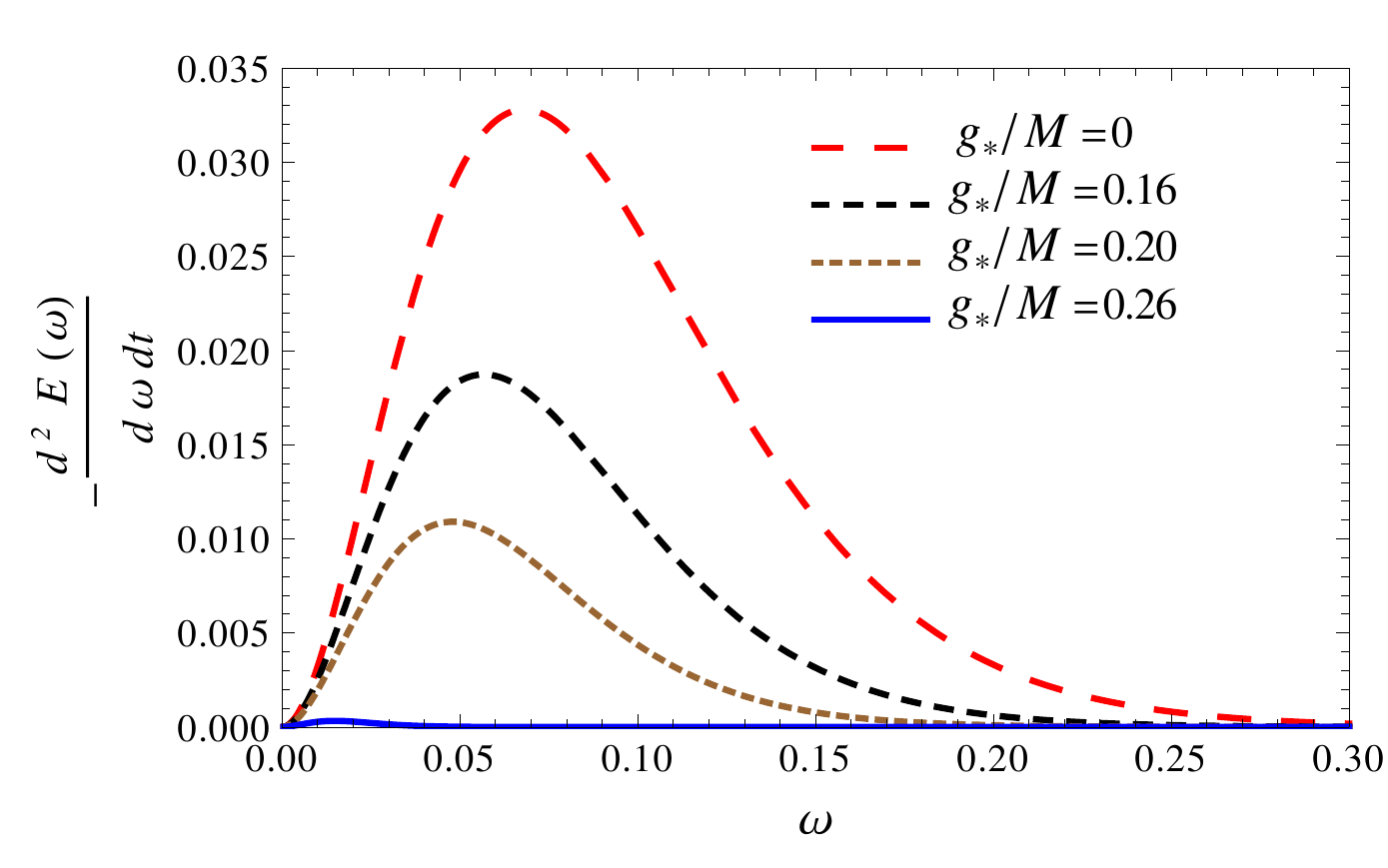}
    \caption{\label{fig8} Plot showing the behavior of energy emission rate versus the frequency ($\omega$) for $a/M=0.9$. (Left)  For Hayward. (Right) For Bardeen.}
\end{figure*}
%%%%%%%%%%

\subsection{Shadow of rotating Hayward and Bardeen black hole}

Now we calculate the celestial coordinates for these black holes which take the following form
\begin{equation}
\alpha=-\xi \csc \theta_{0},
\end{equation}
\begin{equation}
\beta=\pm \sqrt{\eta +a^2 \cos^2 \theta_{0} -\xi^2 \cot^2 \theta_{0}},
\end{equation}
where $\theta_{0}$ is the angle between the rotation axis of the black hole and the line of sight of
the distant observer. {Note that these expressions are valid for a far away observer by the definition (see Eqs. (\ref{defalpha})-(\ref{defbeta})).} If we set the inclination angle to equatorial plane $\theta_{0}=\pi/2$ then the celestial coordinates have the following simple form
\begin{equation}
\alpha=-\xi,
\end{equation}
%%%%%%%%%%%%%%%%%%%%%%%%%%%%%%%%%%%%%%%%
\begin{equation}
\beta=\pm \sqrt{\eta}.
\end{equation}
To visualize the shadow cast by the rotating Hayward and Bardeen black holes, we need to make some plots for the coordinates $\alpha/M$ and $\beta/M$. These plots can be seen from the Figs.~\ref{fig4} and \ref{fig6} for the fixed values of spin parameter $a/M$ and different values of parameters $g/M$ and $g_{*}/M$. We can easily observe the effect of parameters $g/M$ and $g_{*}/M$ on the silhouette of shadow: an increase in the value of $g/M$ and $g_{*}/M$ deceases the size of the silhouette of shadow. The silhouette of shadow is more deformed for the extremal value of $a/M$ (c.f. Figs.~\ref{fig4} and \ref{fig6}).

The behavior of the observable $R_{s}$ and $\delta_{s}$ introduced
in Sect.~\ref{sect2} can be seen from the Figs.~\ref{fig5} and
\ref{fig7}. In both black holes case, we observe that the radius of the silhouette of
shadow decreases and the distortion parameter increases monotonically.

\subsection{Energy emission rate}

Now we plan to discuss the energy emission rate for both rotating Hayward and Bardeen black holes. It has the following form \cite{Wei13}
\begin{equation}
\frac{d^2 E(\omega)}{d\omega dt}= \frac{2 \pi^2 R_{s}^2}{e^{\omega/ T}-1}\omega^3 ,
\end{equation}
where for rotating Hayward black hole the Hawking temperature reads:
\begin{eqnarray}
T_{h} &=& \frac{1}{4 \pi(r_{+}^2+a^2)(r_{+}^3+g^3)}[2r_{+}^2(r_{+}^3+g^3)\nonumber \\
&-&4r_{+}(r_{+}^2+a^2)(r_{+}^3+g^3)+3r_{+}^2(r_{+}^2+a^2)],
\end{eqnarray}
and for rotating Bardeen black hole, it has the following form
\begin{eqnarray}
T_{b} &=& -\frac{1}{4 \pi r_{+}(r_{+}^2+a^2)(r_{+}^2+g_{*}^2)}[4(r_{+}^2+a^2)(r_{+}^2+g_{*}^2) \nonumber \\
&+&2r_{+}^2(r_{+}^2+g_{*}^2)+3r_{+}^2(r_{+}^2+a^2)].
\end{eqnarray}
Next, to see the behavior of the energy emission rate, we plot $d^2 E(\omega)/d\omega dt$ versus $\omega$ for both of the black holes. It can be seen from the Fig.~\ref{fig8} that the representation
is made for spin $a/M=0.9$ and different values of parameters $g/M$
and $g_{*}/M$.

\section{Shadow of regular black hole in the presence of plasma\label{sect4}}

Now we will consider the shadow of the regular black hole in the presence of the plasma. We will use the model of the plasma with the refraction index to be equal to $n=n(x^{i}, \omega)$, where $\omega$ is the photon frequency measured by observer with velocity $u^\alpha $.
So-called effective energy of photon has the form  $\hbar \omega= - p_\alpha  u^\alpha$. In the Ref.~\citep{Synge60}  the expression for the refraction index of the plasma has been obtained in the form:
\begin{equation}
n^2=1+\frac{p_\alpha p^\alpha}{\left( p_\beta u^\beta \right)^2} .
\end{equation}
Note that in the case of absence of the plasma one has the value for the refraction index $n=1$. Using the Hamiltonian for the photon in the form
\begin{equation}
H(x^\alpha, p_\alpha)=\frac{1}{2}\left[ g^{\alpha \beta} p_\alpha p_\beta + (n^2-1)\left( p_\beta u^\beta \right)^2 \right]=0\ ,
\label{generalHamiltonian}
\end{equation}
one can obtain the equations of motion for the photons around regular black holes in the presence of the plasma.
{We introduce the two frequencies of electromagnetic waves, the first one is associated with a timelike Killing vector $\xi^{\alpha}$, i.e., $\omega_{\xi}=-k^{\alpha}\xi_{\alpha}$ and the another one is measured by an observer having a four-velocity $u^{\alpha}$, i.e., $\omega=-k^{\alpha}u_{\alpha}$, where $k^{\alpha}$ is a null wave-vector \cite{Atamurotov15a}
}

Hereafter we will use the  specific form for the plasma frequency in the form
\begin{equation}
n^2=1- \frac{\omega_e^2}{\omega^2},
\label{nFreq}
\end{equation}
where $\omega_e$ is usually called plasma frequency.

Using the Hamilton-Jacobi method described in Sect.~\ref{sect2}, one may easily obtain the equations of motion for the photons around Hayward and Bardeen regular black holes in the presence of plasma as
\begin{eqnarray}
\Sigma\frac{dt}{d\sigma}&=&a ({L_z} - n^2 { E} a
\sin^2\theta)\nonumber\\&&+ \frac{r^2+a^2}{\Delta}\left[(r^2+a^2)n^2 {
E} -a { L_z} \right], \label{teqn}
\\
\Sigma\frac{d\phi}{d\sigma}&=&\left(\frac{{L_z}}{\sin^2\theta}
-a  {E}\right)+\frac{a}{\Delta}\left[(r^2+a^2) { E}
-a { L_z} \right], \label{pheqn}
\\
\Sigma\frac{dr}{d\sigma}&=& \pm \sqrt{\mathcal{R}_p}, \label{reqn}
\\
\Sigma\frac{d\theta}{d\sigma}&=& \pm \sqrt{\Theta_p}, \label{theteqn}
\end{eqnarray}
where the functions $\mathcal{R}_p(r)$ and $\Theta_p(\theta)$ are
introduced as
\begin{eqnarray}
\mathcal{R}_p&=&\left[(r^2+a^2) {E} -a {L_z}
\right]^2+(r^2+a^2)^2(n^2-1){E}^2 \nonumber \\
&&
-\Delta\left[\mathcal{K}+({L_z} -a {E})^2\right]\ , \label{9}
\\
\Theta_p&=&\mathcal{K}+\cos^2\theta\left(a^2  {{E}^2}-\frac{{L_z}^2}{\sin^2\theta}\right) \nonumber\\&& -(n^2-1) a^2 {E}^2 \sin^2\theta\ . \label{10}
\end{eqnarray}
For the plasma frequency $\omega_e$, we will use the expression
\begin{equation}
\omega_e^2=\frac{4 \pi e^2 N(r)}{m_e},
\label{plasmaFreqDef}
\end{equation}
where $e$ and $m_e$ are the electron charge and mass respectively,
and $N(r)$ is the plasma number density.
We will consider a radial power-law density~\cite{Rogers15}
\begin{equation}
N(r)=\frac{N_0}{r^h},
\label{powerLawDensity}
\end{equation}
where $h \geq 0$, such that
\begin{equation}
\omega_e^2=\frac{k}{r^{h}}.
\label{omegaN}
\end{equation}
As an example here we get the value for power $h$ as 1~\cite{Rogers15}.

The shadow of the black hole in the plasma environment can be found using the conditions as it was done in previous sections:
\[{\cal R}(r)=0=\partial {\cal R}(r)/\partial r . \]
Using these equations one can easily find the expressions for the parameters $\xi$ and $\eta$ in the form
\begin{eqnarray}
\xi &=&  \frac{\cal {B}}{{\cal A}} +\sqrt{\frac{{\cal B}^2}{{\cal A}^2} -\frac{{\cal C}}{{\cal A}}}\ , \label{xiexp}\\
\eta&=& \frac{(r^2+a^2-a\xi)^2 +(r^2+a^2)^2 (n^2-1)}{\Delta} \nonumber\\&& -(\xi-a)^2\ , \label{etaexp}
\end{eqnarray}
where we have used the following notations
\begin{eqnarray}
{\cal A}&=& \frac{a^2}{\Delta} \ ,\\
{\cal B}&=& \frac{a}{\Delta}\frac{ ma^2-m r^2+r^3 m'+a^2r m' }{m-r+rm'} \ ,\\
{\cal C}&=& n^2 \frac{(r^2+a^2)^2}{\Delta}\nonumber \\&&+\frac{2r (r^2+a^2)n^2 -(r^2+a^2)^2 n n'}{m-r+rm'} \ ,
\end{eqnarray}
and prime denotes the differentiation with respect to radial coordinate $r$. The functions $\Delta$ and $m$ are defined as in Eqs. (\ref{eqn30})-(\ref{eqn32}).

The expression for the celestial coordinates will take the following form:
\begin{eqnarray}
\alpha&=& -\frac{\xi}{n\sin\theta}\, \label{alpha}\ ,\\
\beta&=&\frac{\sqrt{\eta+a^2-n^2a^2\sin^2\theta-\xi^2\cot^2\theta }}{n} \label{beta}\ ,
\end{eqnarray}
for the case when black hole is surrounded by plasma.

\begin{figure*}[t!]
\begin{center}

\includegraphics[width=0.24\linewidth]{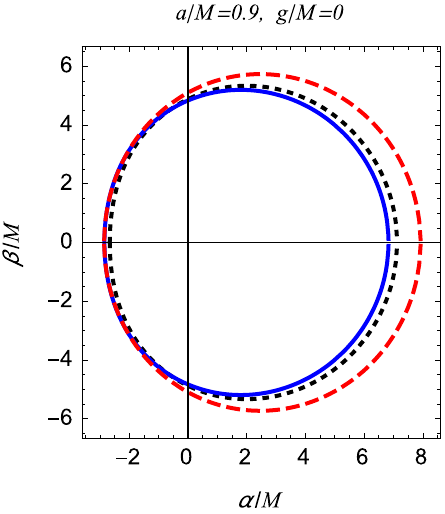}
\includegraphics[width=0.24\linewidth]{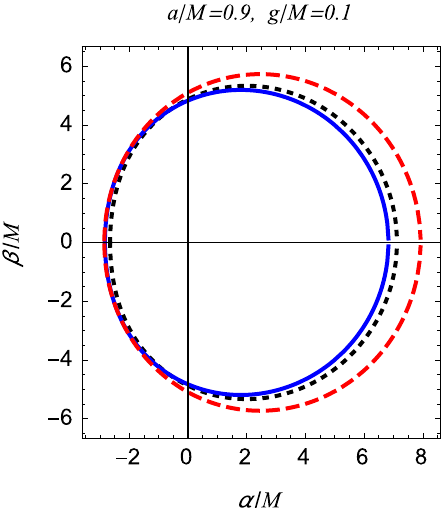}
\includegraphics[width=0.24\linewidth]{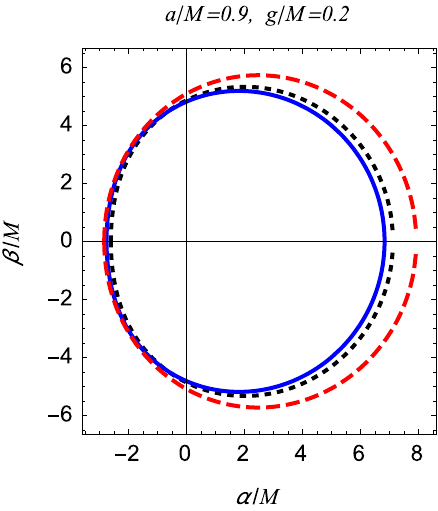}
\includegraphics[width=0.24\linewidth]{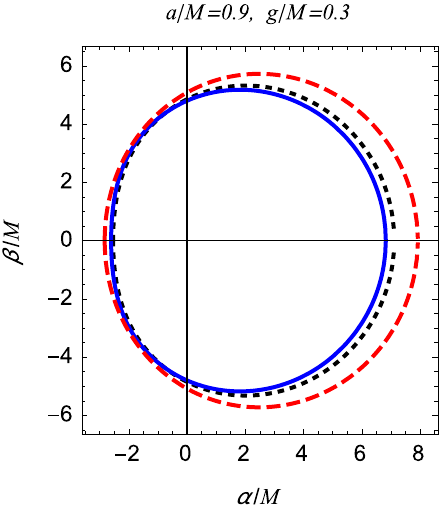}

\includegraphics[width=0.24\linewidth]{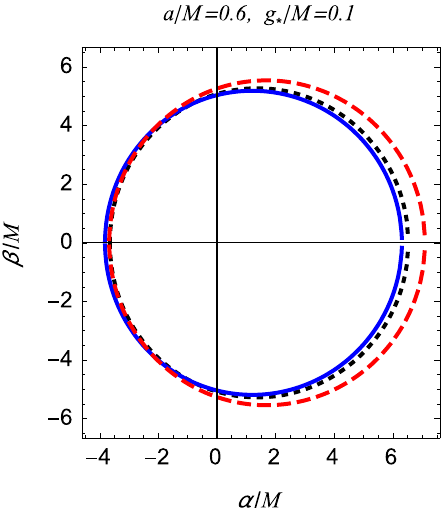}
\includegraphics[width=0.24\linewidth]{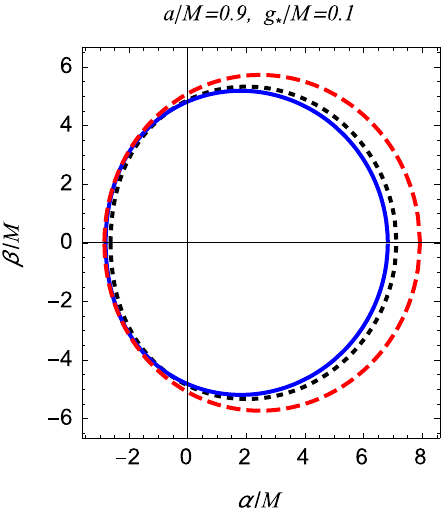}
\includegraphics[width=0.24\linewidth]{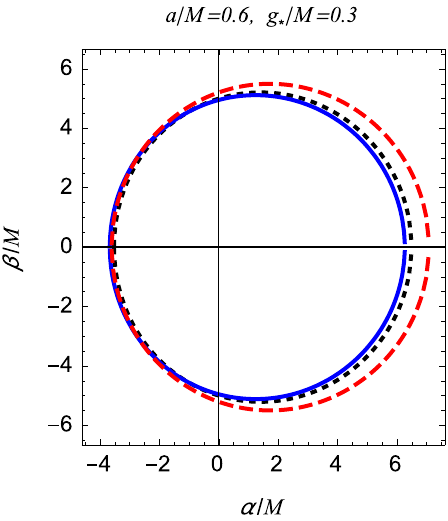}
\includegraphics[width=0.24\linewidth]{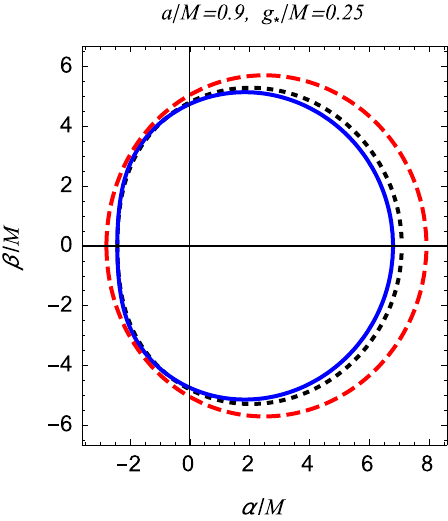}

\end{center}
\caption{Plot showing the silhouette of the shadow cast by the Hayward and Bardeen regular black holes surrounded by plasma for the different values of the rotation parameter $a/M$, and the refraction index. The solid lines in the plots correspond to the vacuum case, while for dotted and dashed lines we choose the plasma frequency $\omega_e/\omega_{\xi}=k/r$, where $(k/M)^2 = 0.5$ and $(k/M)^2 = 1.0$, respectively. The inclination angle between observer and the axis of the rotation has been taken to be $\theta_0=\pi/2$\label{fig10}}
\end{figure*}

In Fig.~\ref{fig10} the silhouette of shadow of the rotating
Hayward and Bardeen black holes for the different values of black
hole rotation parameter $a/M$, parameters $g/M$ and $g_{*}/M$ have been presented.
In these figures we choose the plasma frequency in the form $\omega_e/\omega_\xi=k/r$.
From the Fig.~\ref{fig10} one can easily see that the presence of the plasma affects the apparent size of the shadow to be increased while we have shown that the parameters of the regular black holes force to decrease the shadows size. There is also tendency of decreasing the distortion of the shadow in the presence of plasma. Physically this is similar as an effect of gravitational redshift of photons in the gravitational field of the regular black holes: the frequency change due to gravitational redshift affects the plasma refraction index.

\section{Conclusion\label{sectconcl}}

In this paper, we have analyzed the shape of the shadow cast by
different types of regular black holes. We have discussed how the
shadow cast by these black holes is distorted by the presence of the
various parameters related to regular black holes and environment.
In first part of the paper we have studied the shadow cast by the
rotating ABG black hole. We have found that the presence of electric
charge affected the shape of the shadow. We see that with increasing
the value of $a/M$ the shape of the shadow becomes more and more
asymmetric with respect to the vertical axis. It can be seen that
for the fixed value of spin the size of the shadow is monotonically
decreases as the electric charge increases. Furthermore, we calculate
the deformation due to an increase in the spin of black hole which
is characterized by the deformation parameter ($\delta_{s}$). One
can see that $\delta_{s}$ increases with an increase in the electric
charge as well as spin. Next, we have discussed about the energy
emission rate of the rotating ABG black hole. The energy emission
rate decreases with an increase in the value of $Q$ as well as in
$a/M$. It can be seen from the Fig.~\ref{fig3} that the peak is
sharp for small values of $Q$.

In the next part  of the paper we have studied the shadow of the rotating Hayward and Bardeen black holes. The rotating Hayward black hole contains $g/M$ which provides deviation from the Kerr black hole and the rotating Bardeen black hole has another parameter $g_{*}/M$ which is a magnetic charge due to the non-linear electromagnetic field. Furthermore, we see the effect of the parameters $g/M$ and $g_{*}/M$ on the shape of black hole shadow. We have found that the presence of the parameters $g/M$ and $g_{*}/M$ decreases the size of the silhouette of shadow for each fixed value of $a/M$. There is an increase in $\delta_{s}$ when $a$ is increasing and get an extremal value for both black holes. One can also see the behavior of energy emission rate versus frequency which indicates that the energy emission rate decreases with increasing value of $g/M$ and $g_{*}/M$.

In the last part of the paper we have also studied the influence of the plasma environment around Hayward and Bardeen regular black holes to the change of the size and shape of the regular black hole's shadow. It was shown that the presence of the plasma affects the apparent size of the shadow to be increased while we observe opposite effect by magnetic charges of the regular black holes. There is also a tendency of decreasing the distortion of the shadow in the presence of plasma. Physically this is similar as an effect of gravitational redshift of photons in the gravitational field of the regular black holes: the frequency change due to gravitational redshift affects the plasma refraction index.

\begin{acknowledgements}
M.A. acknowledges the University Grant Commission, India, for financial support through the Maulana Azad
National Fellowship For Minority Students scheme (Grant No.~F1-17.1/2012-13/MANF-2012-13-MUS-RAJ-8679) and to the Institute of Nuclear Physics, Uzbekistan Academy of Sciences for a great hospitality where this work was done. The research is supported in part by Projects  No.~F2-FA-F113, No.~EF2-FA-0-12477, and No.~F2-FA-F029 of the UzAS,
and by the ICTP through Grants No.~OEA-PRJ-29 and No.~OEA-NET-76 and by the Volkswagen Stiftung, Grant No.~86 866. SGG would like to thank SERB-DST Research Project Grant NO SB/S2/HEP-008/2014
\end{acknowledgements}

\bibliographystyle{apsrev4-1}  %% BibTeX style
\bibliography{/hp/ahmadjon_hp/Nauka/gravreferences/gravreferences}

\end{document}